\documentclass[hidelinks,11pt]{article}
\usepackage{amsmath}
\usepackage{amsfonts}
\usepackage{color,soul}
\usepackage[table]{xcolor}
\usepackage{natbib}
\usepackage{threeparttable} 
\usepackage{hyperref}
\usepackage{booktabs}
\usepackage{graphicx}
\usepackage{epstopdf}
\usepackage{lipsum}
\usepackage{cleveref}
\usepackage{rotating}
\usepackage{caption}
\usepackage{subcaption}
\usepackage{multirow}
\usepackage[stable]{footmisc}
\usepackage{breqn}
\usepackage{booktabs}
\usepackage{rotating}
\usepackage{floatrow}
\usepackage{mathtools}
\usepackage{breqn}

\title{Measurement of Common Risk Factors: A Panel Quantile Regression Model for Returns
\thanks{The support from the Czech Science Foundation under the 16-14151S project and the support from the Grant Agency of Charles University under the 610317 project is gratefully acknowledged. This project has received funding from the European Union’s Horizon 2020 Research and Innovation Staff Exchange programme under the Marie Sklodowska-Curie grant agreement No 681228. We would also like to thank James L. Powell, Antonio F. Galvao and seminar participants at 3rd International Workshop on Financial Markets and Nonlinear Dynamics in Paris (FMND 2017) as well as  conference participants at the 2015 CFE Network Conference in London, the Joint Annual Meeting of the Slovak Economic Association and the Austrian Economic Association (NOeG-SEA 2016),  the 2016 CFE Network Conference in Seville and the 2017 IAAE meetings in Sapporo for helpful comments.}
}

\author{Franti\v{s}ek \v{C}ech\thanks{Institute of Economic Studies, Charles University, Opletalova 26, 110 00, Prague,  CR and Institute of Information Theory and Automation, Academy of Sciences of the Czech Republic, Pod Vodarenskou Vezi 4, 182 00, Prague, Czech Republic.Phone: +420 776 535 106 Email: frantisek.cech@fsv.cuni.cz} \and Jozef Barun\'{\i}k\thanks{Institute of Economic Studies, Charles University, Opletalova 26, 110 00, Prague,  CR and Institute of Information Theory and Automation, Academy of Sciences of the Czech Republic, Pod Vodarenskou Vezi 4, 182 00, Prague, Czech Republic. Phone: +420 776 259 273. Email: barunik@utia.cas.cz}} 

\begin{document}
\maketitle

\begin{abstract}
\noindent This paper investigates how to measure common market risk factors using newly proposed Panel Quantile Regression Model for Returns. By exploring the fact that volatility crosses all quantiles of the return distribution and using penalized fixed effects estimator we are able to control for otherwise unobserved heterogeneity among financial assets. Direct benefits of the proposed approach are revealed in the portfolio Value--at--Risk forecasting application, where our modeling strategy performs significantly better than several benchmark models according to both statistical and economic comparison. In particular Panel Quantile Regression Model for Returns consistently outperforms all the competitors in the 5\% and 10\% quantiles. Sound statistical performance translates directly into economic gains which is demonstrated in the Global Minimum Value--at--Risk Portfolio and Markowitz-like comparison. Overall results of our research are important for correct identification of the sources of systemic risk, and are particularly attractive for high dimensional applications.\\

\noindent \textbf{JEL Classification}: C14, C23, G17, G32 \\
\noindent \textbf{Keywords}: panel quantile regression, realized measures, Value--at--Risk
\end{abstract}

\newpage
\section{Introduction}

Many studies document cross-sectional relations between risk and expected returns, generally measuring a stock's risk as the covariance between its return and some factor. In this laborious search for proper risk factors,\footnote{See for example \cite{harvey2016and,feng2017taming} for recent very complete overviews. This research dates back to \cite{french1987expected}.} volatility still plays central role in explaining expected stock returns for decades.  Most recent efforts explore increasingly available datasets, and make measurement of ex-post volatility more precise than ever before. In turn, these measures can be used for more precise identification of market risk. Although predictions about expected returns are essential for understating of classical asset pricing, little is known about potential of the factors to precisely identify extreme tail events of the returns distribution. More importantly, even less is known about commonalities between more assets with this respect. Our research attempts to contribute in this direction. 

Asset pricing models explaining risk valuation theoretically assume an economic agent who decides based on the preference about her consumption by maximizing expected utility function. However, these preferences may be too restrictive to deliver satisfactory description of the real behavior of agents. Instead of working with standard expected utilities, recent literature strives to incorporate heterogeneity into dynamic economic models assuming agents maximize their stream of future quantile utilities \citep{chambers2007ordinal,rostek2010quantile,de2017dynamic}. We contribute to these efforts by developing a Panel Quantile
Regression Model for Returns that is able to control for otherwise unobserved heterogeneity among financial assets and allows us to exploit common factors in volatility that directly affect future quantiles of returns. In a sense, we revisit large literature connecting volatility with cross-section of returns, as by construction, we model tail events of the conditional distributions via volatility.

Since the seminal work of \cite{koenker1978regression}, quantile regression models have been increasingly used in many disciplines. In finance, \cite{engle2004caviar} were among the first to use quantile regression to develop the Conditional Autoregressive Value--at--Risk (CAViaR) model and capture conditional quantiles of the asset returns. \cite{baur2012stock} use quantile autoregressions to study conditional return distributions, \cite{cappiello2014measuring} detects comovements between random variables with time-varying quantile regression. \cite{Zikes_Barunik2015semi} show that various realized measures are useful in forecasting quantiles of future returns without making assumptions about underlying conditional distributions. Resulting semi-parametric modeling strategy captures conditional quantiles of financial returns well in a flexible framework of quantile regression. Moving the research focus towards multivariate framework, and concentrating on interrelations between quantiles of more assets, \cite{white2015var} pioneers the extension. Different stream of multivariate quantile regression based literature concentrates on the analysis using factors \citep{chen2016quantile,ando2017quantile}.\footnote{Panel quantiles methods are useful in the other areas of economics besides finance. They are mostly applied in the labour economics (\cite{billger2015panel}, \cite{dahl2013wage}, \cite{toomet2011learn}), banking and economic policy analysis (\cite{covas2014stress},  \cite{klomp2012banking}), economics of education  (\cite{lamarche2008private}, \cite{lamarche2011measuring}), energy and environmental economics (\cite{you2015democracy}, \cite{zhang2015direct}) or international trade (\cite{dufrenot2010trade}, \cite{foster2014importing}, \cite{powell2014exporter}).} From the theoretical point of view, \cite{giovannetti2013asset} derives an asset pricing model in which equity premium is no longer based on the covariance between return and consumption. Instead, \cite{giovannetti2013asset} argue that under optimism, higher volatility can be connected to high chance of high returns leading to increased prices, hence decreasing expected returns, and vice versa under pessimism. Based on Choquet utility functions, \cite{bassett2004pessimistic} show that pessimistic optimization may be formulated as a linear quantile regression problem, and can lead to optimal portfolio allocation. 

With this respect, work by \cite{Zikes_Barunik2015semi} is important as it provides link between future quantiles of return distribution and its past variation. As the financial sector is highly connected and the co-movements in asset prices are common, there is a need for proper identification of dependencies in joint distributions. In the classical mean-regression framework, \cite{bollerslev2016risk} showed that realized volatility of financial time series shares many commonalities. In the quantile regression set-up, however, there is no similar study that will try to uncover information captured in the panels of volatility series.   Moreover, to the best of our knowledge there is no study estimating conditional distribution of returns in a multivariate setting that explores ex-post information in the volatility. 

In this paper, we contribute to the literature by introducing Panel Quantile Regression Model for Returns. Our model utilize all the advantages offered by panel quantile regression and financial market datasets. In particular, we are able to control for otherwise unobserved heterogeneity among financial assets and reveal common factors in volatility that have direct influence on the future quantiles of returns. To the best of our knowledge this is  one of the first applications of the panel quantile regression using dataset where time dimension $T$ is much greater than cross-sectional dimension $N$, i.e. $T>>N$. As a result we are able to obtain estimates of quantile specific individual fixed effects that represents idiosyncratic part of the market risk. 

In an empirical application, we hypothesize that newly proposed model will deliver more accurate estimates compared to currently established methods. These estimates moreover translates into better forecasting performance of Panel Quantile Regression Model for Returns. In addition, using penalized fixed effect estimator we will be able to disentangle overall market risk into systematic and idiosyncratic parts. Actual performance of our model is tested in the portfolio Value--at--Risk forecasting exercise. Before the analysis of the empirical dataset (29 highly liquid stocks from the New York Stock Exchange) we run small Monte-Carlo experiment that enable us to study well-behaved data. For the robustness reasons we evaluate forecasts from both statistical and economic perspective. In the statistical comparison we furthermore distinguish between absolute and relative performance of the given model. 

Results of our analysis suggest that the Panel Quantile Regression Model for Returns is dynamically correctly specified. Moreover it dominates the benchmark models in the economically important quantiles (5\%,10\% or 95\%). Overall we find that according to statistical comparison none of the benchmark models is able to outperform our model consistently. Furthermore model we introduce in this paper provide us with direct economic gains according to both economic evaluation criteria.

%

\section{Risk Measurement using High Frequency Data} \label{sec:theory}

Let's assume that the efficient logarithmic price process $p_{i,t}$ of $i$th asset evolves over time $0\le t \le T$ according to the following dynamics
\begin{equation}
dp_{i,t} = \mu_{i,t} dt + \sigma_{i,t} dW_{i,t} + dJ_{i,t},
\end{equation} 
where $\mu_{i,t}$ is a predictable component, $\sigma_{i,t}$ is cadlag process, $W_{i,t}$ is a standard Brownian motion, and $J_{i,t}$ is a jump process. 

The volatility of the logarithmic price process can be measured by quadratic return variation which can be decomposed into integrated variance (IV) of the price process and the jump variation (JV): 
\begin{equation}
QV_{i,t}=\underbrace{\int_{t-1}^t \sigma_{i,s}^2ds}_{IV_{i,t}} + \underbrace{\sum_{l=1}^{N_{i,t}} \kappa_{i,t,l}^2}_{JV_{i,t}},
\end{equation}
where $N_{i,t}$ is total number of jumps during day $t$ and $\sum_{l=1}^{N_t} \kappa_{i,t,l}^2$ represents magnitude of the jumps. As shown by \cite{Andersen2003} Realized Variance estimator can be simply constructed by squaring intraday returns: 
\begin{equation}
\widehat{RV}_{i,t}=\sum_{k=1}^N \left(\Delta_k p_{i,t} \right)^2,
\end{equation}  
where $\Delta_k p_{i,t} = p_{i,t-1+\nu_k/N}-p_{i,t-1+\nu_{k-1}/N}$ is a discretely sampled vector of $k$-th intraday log-returns of $i$th asset in $[t-1,t]$, with $N$ intraday observations. Realized Variance estimator moreover converges uniformly in probability to $QV_{i,t}$ as the sampling frequency goes to infinity $$\widehat{RV}_{i,t} \xrightarrow[N\rightarrow \infty]{p} \int_{t-1}^t \sigma_{i,s}^2ds + \sum_{l=1}^{N_t} \kappa_{i,t,l}^2 $$

Building on the concept of Realized Variance \cite{barndorff2004power} and \cite{barndorff2006econometrics} introduced bipower variation estimator that is robust to jumps and thus able to consistently estimate $IV_{i,t}$. Furthermore, \cite{andersen2011reduced} adjust original estimator, which helps render it robust to certain type of microstructure noise: $$\widehat{IV}_{i,t}^{BPV}=\mu_{1}^{-2} \left(\frac{N}{N-2}\right)\sum_{k=3}^N \vert \Delta_{k-2} p_{i,t} \vert  \vert \Delta_{k} p_{i,t} \vert ,$$ where $\mu_{\alpha}=E(\vert Z^\alpha \vert)$, and $Z \sim N(0,1)$. Having estimator of $IV_{i,t}$ in hand jump variation can be consistently estimated\footnote{Asymptotic behaviour and further details of the estimator can be found in \cite{barndorff2006econometrics}.} as a difference between Realized Variance and the bipower variation: $$\left( \widehat{RV}_{i,t} - \widehat{IV}_{i,t}^{BPV}\right) \xrightarrow[N\rightarrow \infty]{p} \sum_{l=1}^{N_t} \kappa_{i,t,l}^2.$$  
 
For many financial applications not only magnitude of the variation but also its sign is important. Therefore \cite{barndorff2010measuring} introduce innovative approach for measuring negative and positive variation in data called Realized Semivariance. They showed that Realized Variance can be decomposed to realized downside semivariance ($RS_{i,t}^-$) and realized upside semivariance ($RS_{i,t}^+$): $$RV_{i,t} = RS_{i,t}^+ + RS_{i,t}^-,$$
where $RS_{i,t}^+$ and $RS_{i,t}^-$ are defined as follows, 
\begin{equation}
\widehat{RS}^+_{i,t}=\sum_{k=1}^N \left(\Delta_k p_{i,t} \right)^2I\left(\Delta_k p_{i,t}  > 0\right)\xrightarrow[]{p} \frac{1}{2}IV_{i,t} +  \sum_{l=1}^{N_t} \kappa_{i,t,l}^2 I \left(\kappa_{i,t,l} > 0\right)
\end{equation}     
\begin{equation}
\widehat{RS}^-_{i,t}=\sum_{k=1}^N \left(\Delta_k p_{i,t}\right)^2I\left(\Delta_k p_{i,t}  < 0\right)\xrightarrow[]{p} \frac{1}{2}IV_{i,t} +  \sum_{l=1}^{N_t} \kappa_{i,t,l}^2 I \left(\kappa_{i,t,l} < 0\right).
\end{equation}

Consequently, the negative and positive semivariance provides information about variation associated with movements in the tails of the underlying variable. Similarly to \cite{patton2015good,bollerslev2016good}, we use negative semivariance as a proxy to the bad state of the returns, and positive semivariance as an empirical proxy of the good state of the underlying variable.

Since correlation is inevitably important in portfolio applications, and we use it later in portfolio Value--at--Risk application, we also define Realized Covariance estimator \citep{Barndorff-Nielsen2004} as 
$$
\widehat{\Sigma}_{t}=\sum_{k=1}^N \mathbf{\left(\Delta_k p_t \right) \left(\Delta_k p_t\right)'},
$$ where $\mathbf{\Delta_k p_t}=(\Delta_k p_{1,t},...,\Delta_k p_{q,t})'$ is vector containing log-returns of $q$ individual assets.

\section{Panel Quantile Regression Model for Returns}

Having briefly described realized measures that we need for model construction, we now propose simple linear models for cross-section of quantiles of  future returns. We base our model in a recent theoretical endeavors to move from expected values to quantiles and understand heterogeneity in asset prices. Based on the risk preferences of quantile maximizers \citep{manski1988ordinal,rostek2010quantile}, \cite{de2017dynamic} develop a dynamic model of rational behavior under uncertainty, in which agent maximizes stream of future quantile utilities. This is in sharp contrast to the mainstream literature that assumes decision making process to be driven by maximization of the expected utility instead. In a similar spirit as in \cite{bassett2004pessimistic}, our model can be viewed as linear asset pricing equation
\begin{equation}\label{eq:quantile dynamics}
Q_{r_{i,t+1}}(\tau\lvert v_{i,t})=\alpha_i(\tau)+v_{i,t}^{\top}\beta(\tau),  \quad \tau \in (0,1),
\end{equation} 
where $r_{i,t+1}=p_{i,t+1}-p_{i,t}$ are logarithmic daily returns, $v_{i,t}=\Big(\widehat{QV}_{i,t}^{1/2},\widehat{QV}_{i,t-1}^{1/2},\dotsc,\widehat{IV}_{i,t}^{1/2},$ $\widehat{IV}_{i,t-1}^{1/2},\dotsc,\widehat{JV}_{i,t}^{1/2},\widehat{JV}_{i,t-1}^{1/2},\dotsc \Big)$ are individual components of the quadratic variation, $\alpha_i$ represents individual fixed effects. This model enables us to study influence of the individual fixed effects $\alpha_i$ and coefficient estimates $\beta$ on the specific quantiles of the future returns. \autoref{eq:quantile dynamics} can be easily extended by exogenous variables such as factors used in \cite{fama1993common}, as already attempted by \cite{galvao2017testing}.

To obtain the parameters defined in \autoref{eq:quantile dynamics} we use panel quantile regression as introduced in \cite{koenker2004quantile}. In this seminal work Roger Koenker proposed a penalized fixed effects estimator as a general approach to estimating quantile regression models in the panel data framework. Recently the ideas of \cite{koenker2004quantile} have been further developed  by \cite{lamarche2010robust} who studied penalized quantile regression estimator, \cite{galvao2011quantile} where the fixed effects model for dynamic panels is introduced, \cite{galvao2010penalized} where it is shown that bias in dynamic panels can be reduced using penalty term, work of \cite{canay2011simple} who introduced simple two-step approach to estimation of panel quantile regression and showed consistency and asymptotic normality of the proposed estimator, or application of the instrumental variables to quantile regression estimation \citep{harding2009quantile}. Other influential works developing theory of panel quantile methods are \cite{galvao2015bootstrap}, \cite{galvao2015efficient}, \cite{galvao2015smoothed},  \cite{Graham2015panel}, \cite{harding2014estimating} or \cite{kato2012asymptotics}.

In our work we propose to model quantiles of several return series using original penalized fixed effects estimator of \cite{koenker2004quantile}. The advantage of this approach is ability to account and to control for unobserved heterogeneity among financial assets which will yield more precise quantile specific estimates. As a consequence these estimates will translate into better forecasting performance directly. Moreover one can use this approach to obtain precise estimates of the Value--at--Risk (VaR) which is commonly used financial industry risk measure. In the VaR application panel data will utilize all the favorable properties of the standard time series. In additon, cross-sectional dimension will help us to account for common shocks among the assets. To obtain parameter estimates we solve following optimization problem
\begin{equation} \label{eq:PQR equation}
\min\limits_{\alpha(\tau),\beta(\tau)}\sum_{t=1}^{n}\sum_{i=1}^{t_i} \rho_{\tau}(r_{i,t+1}-\alpha_i(\tau)-v_{i,t}^{\top}\beta(\tau))  + \lambda\sum_{i=1}^n \lvert \alpha_i(\tau)\rvert,
\end{equation}
where $\rho_{\tau}(u)=u\left(\tau-I(u(<0))\right)$ is the quantile loss function \citep{koenker1978regression} and $\sum_{i=1}^n \lvert \alpha_i\rvert$ is $l_1$ penalty that controls variability introduced by the large number of estimated parameters. In our set-up individual fixed effects are consider to have distributional effects and we concentrate on each quantile separately rather than minimizing through several quantiles. In contrast, \cite{koenker2004quantile} and vast majority of the theoretical and applied works consider $\alpha_i$ to have a pure location shift effect on the conditional  quantiles. This restriction is a consequence of the structure of the usual panel-datasets where cross-sectional dimension is much larger than time dimension\footnote{As detailed in \cite{koenker2004quantile} it is not advisable to estimate $\tau$-specific $\alpha_i$ in problems with small/medium $T$.}. This problem is however not so sever in analysis of financial market data because majority of the assets have long history and thus consist of thousands of observations. Moreover analysis of the specific quantiles is essential for many financial applications including popular Value--at--Risk in which we are most often interested in finding 1-day 5\% VaR or 10-day 1\% VaR as historically recommended by Basel Committee on Banking Supervision.

Another important part of the \autoref{eq:PQR equation} besides fixed effects is the penalty term $\lambda$ which influence the precision of the estimates of $\alpha_i(\tau)$ and $\beta(\tau)$. Our analysis starts with the standard pure fixed effects model where $\lambda=0$. This approach allows us to obtain estimates of all individual quantile specific fixed effects. As a robustness check we also carried out the analysis with values of $\lambda$ from range $(0;1)$ as in \cite{damette2012economic} and \cite{covas2014stress} and with $\lambda=1$ as in \cite{koenker2004quantile}, \cite{bache2008determinants}, \cite{matano2011wage}, \cite{lee2012momentum} and \cite{you2015democracy}. Overall we find out that choice of $\lambda$ does not affect precision of $\beta$ estimates. We address this finding to the structure and characteristics of the dataset (high time dimension $T$ compared to low cross-section dimension $N$). Although parameter $\lambda$ is set arbitrarily in our work there exist also theoretical approach of $\lambda$ selection. Interested reader can find it in \cite{lamarche2010robust} or \cite{galvao2010penalized}.

\subsection{Model specifications}

While \autoref{eq:quantile dynamics} can accommodate many possible model specifications, we are interested in estimation results for following three models. In each specification quantiles of return series depends on different realized measure - Realized Volatility, Realized Semivariances and Realized Bi-Power Variation. Hence, we are estimating following Panel Quantile Regression Models for Returns specifications: 
\begin{enumerate}
\item \textit{PQR-RV} for Realized Volatility, with quantile function defined as 
\begin{equation}
Q_{r_{i,t+1}}(\tau)=\alpha_i(\tau) + \beta_{RV^{1/2}}(\tau)*RV_{t}^{1/2},
\end{equation} 
\item \textit{PQR-RSV} for Realized Semivariance, with quantile function defined as 
\begin{equation}
Q_{r_{i,t+1}}(\tau)=\alpha_i(\tau) + \beta_{{RS^+}^{1/2}}(\tau)*{RS_t^+}^{1/2}+ \beta_{{RS^-}^{1/2}}(\tau)*{RS_t^-}^{1/2},
\end{equation} 
\item \textit{PQR-BPV} for Realized Bi-Power Variation, with quantile function defined as 
\begin{equation}
Q_{r_{i,t+1}}(\tau)=\alpha_i(\tau) +\beta_{BPV^{1/2}}(\tau)*BPV_t^{1/2}+ \beta_{Jumps^{1/2}}(\tau)*Jumps_t^{1/2}.
\end{equation} 
\end{enumerate}

\section{Competing Models and Evaluation }
In the previous section we introduce Panel Quantile Regression Model for Returns which will be used in the applied part of the paper to analyze simulated and empirical data. In this section we describe alternative approaches that can be viewed as the direct competitors to our model. Benchmarks in our work includes popular and widely used RiskMetrics model that is the industry standard for the risk evaluation in high-dimensional problems and two applications of the Univariate Quantile Regression Model for Returns.
\subsection{RiskMetrics}
Based on Exponentially Weighted Moving Average, J.P. Morgan Chase in 1996 introduced new methodology for accessing the financial risk called RiskMetrics. It is considered to be the baseline benchmark model for numerous fiancial applications.  For our benchmark purposes, we adopt the specification in its original form as defined in \cite{Longerstaey1996} with decay factor, $\lambda$ set to 0.94. We assume a $q\times 1$ vector of daily returns  $r_t=\sum_{k=1}^n \left(\Delta_k p_t \right)$ for $t=1,...,T$ such that $r_t \sim N\left(\mu_t,\sigma_t^2 \right)$, where $\mu_t$ is conditional mean and $\sigma_t^2$ is conditional variance of daily returns. We also assume that $\mu_t=0$ and therefore conditional covariance has the form 
$$\sigma_{i,j,t}=\lambda\sigma_{i,j,t-1}+(1-\lambda)r_{i,t-1}r_{j,t-1},$$
where $\sigma_{i,j,t}$ denotes covariance between assets $i$ and $j$ at time $t$. 
\subsection{Univariate Quantile Regression Model for Returns}
As already mentioned \cite{Zikes_Barunik2015semi} introduced elegant framework for modelling and obtaining forecasts of the conditional quantiles of future returns in the univariate setting. They proposed to model quantiles of return series according to:
\begin{equation}\label{univariate QR}
Q_{r_{i,t+1}}(\tau\lvert v_{i,t})=\alpha_i(\tau)+v_{i,t}^{\top}\beta_i(\tau),
\end{equation} 
where $r_{i,t+1}=p_{i,t+1}-p_{i,t}$ is return series of $i$th asset and $v_{i,t}=\Big(\widehat{QV}_{i,t}^{1/2},\widehat{QV}_{i,t-1}^{1/2},\dotsc,\widehat{IV}_{i,t}^{1/2},$ $\widehat{IV}_{i,t-1}^{1/2},\dotsc ,\widehat{JV}_{i,t}^{1/2},\widehat{JV}_{i,t-1}^{1/2},...\Big)$  are components of quadratic variation. Estimates of asset $i$ quantile specific $\beta$ from \autoref{univariate QR} are obtained by minimizing following objective function:
\begin{equation}
\min\limits_{\alpha_i(\tau),\beta_i(\tau)}\frac{1}{n}\sum_{t=1}^{n}\rho_{\tau}\left(r_{i,t+1}-\alpha_i(\tau)-v_{i,t}^{\top}\beta_i(\tau)\right),
\end{equation} 
where $\rho_{\tau}(u)=u(\tau-I(u<0)))$ is the quantile loss function defined in \cite{koenker1978regression}. Application of the model in the multivariate setting is further described in the following section.

\subsection{Forecasting Exercise and Forecast Evaluation}
In order to evaluate performance of the newly proposed Panel Quantile Regression Model for Returns we conduct forecasting exercise in which we study portfolio Value--at--Risk from the statistical and economic point of view. We decided to concentrate on both statistical and economic evaluation in order to get complete picture of behavior of the new model. Moreover concentrating on statistical evaluation only might get us into trouble because good statistical performance might not necessarily translates also into economic gains. Therefore to make our results robust we apply two statistical and two economic evaluation criteria.    

In statistical comparison we focus on the absolute and relative performance of the considered models in the equally weighted portfolio set-up. By focusing on the equally weighted portfolio, we refrain from specifying complicated weighting scheme which might theoretically affects the overall performance. 

In economic comparison we study efficient frontier of the Value--at--Risk - return trade-off and also Global Minimum Value--at--Risk Portfolio (GMVaRP). As both approaches by definition tries to find optimal weights of the assets we are not using equally weighted portfolio here anymore. 
\subsubsection{Portfolio Value--at--Risk}
Value--at--Risk is elegant way of quantifying risk of an investment. Its simplicity makes it popular in the financial industry because it provides us with single number that represents potential loss we can incur at certain probability level during pre-defined period of time. Use of VaR as the only risk measure however has some limitations. There are well known problems of  VaR generally not being a coherent risk measure because of violating the subadditivity criteria \citep{artzner1999coherent}. However, \cite{danielsson2013fat} show that under reasonable assumptions VaR might be subadditive. In this paper we decided to use VaR framework because forecasts we obtain from the Panel Quantile Regression Model for Returns are by definition semi-parametric VaRs. Moreover we are not trying to introduce new measure of financial risk, rather we want to show accuracy of the model we proposed in the standard well--known set--up.

Having briefly discussed our motivation to concentrate on the VaR in our analysis we now turn to Value--at--Risk framework itself. Generally there are two main approaches of calculating VaR: (semi)parametric estimation vs. historical simulation. In our work we will concentrate on parametric approach because it directly enable us to easily compare forecasts from several benchmark models.

Original parametric way of VaR calculation was introduced by J.P.Morgan. In their set-up, VaR is derived from quantile of standard normal distribution, $$VaR_i=\gamma_\tau \sigma_i,$$ where $\gamma_\tau$ is the $\tau$ quantile of the standard normal distribution and $\sigma_i$ is the volatility of the asset $i$. If we would like to study VaR of the portfolio instead of the individual assets, $\sigma_i$ is replaced by the portfolio volatility $\sigma_P$. Under assumption of the multivariate normality $\sigma_P$ is calculated as $$\sigma_{P}=\sqrt{w^{\top}*\Sigma*w},$$ where $\Sigma$ is the covariance matrix and $w$ is the vector of asset weights. We can therefore calculate percentage Value--at--Risk ($\%VaR$) of the given portfolio as a  
\begin{equation}
\label{standard_VaR}
\%VaR_P=\sqrt{\gamma_\tau^2*w^{\top}*\Sigma*w}.
\end{equation}

We can rewrite \autoref{standard_VaR} in terms of VaRs of the individual assets as 
\begin{equation}
\label{matrix_VaR}
\%VaR_P=\sqrt{(w^{\top}\odot\%VaR^{\top})*\Omega*(w\odot\%VaR)},
\end{equation}
where $\%VaR$ is a vector of individual percentage VaR estimates, $\Omega$ stands for correlation matrix and $\odot$ is the Hadamar product. Alternatively we can also write it as $$\%VaR_P=\sqrt{\sum_{i=1}^N{(w_i\%VaR_i)}^2+2\sum_{i=1}^N\sum_{j=i+1}^Nw_iw_j\%VaR_i\%VaR_j\rho_{i,j}}$$
where $w_i$ is the weight of asset $i$, $\%VaR_i$ is the percentage VaR of the $i^{th}$ asset and $\rho_{i,j}$ represents correlation between asset $i$ and $j$. 

In the forecasting exercise we will study portfolio Value-at-Risk performance of the 4 benchmark model specifications: 
\begin{itemize}
\item RiskMetrics,
\item Panel Quantile Regression(PQR) Model for Returns,
\item Univariate Quantile Regression(UQR) Model for Returns,
\item portfolio version of Univariate Quantile Regression(Portfolio UQR) Model for Returns.
\end{itemize}
For calculation of portfolio VaR using \textit{RiskMetrics} approach we directly apply \autoref{standard_VaR} where $\Sigma$ is covariance matrix obtained from RiskMetrics and $\gamma_\tau$ is a cut--off point of standard normal distribution at a given quantile $\tau$.

In case of \textit{PQR} and \textit{UQR} forecasts of quantiles of return series are considered to be semi-parametric percentage VaR. Correlation matrix $\Omega$ is obtained from Realized Covariance matrix estimate, $\Sigma$, as $$\Omega=\left(diag(\Sigma)\right)^{-1/2}*\Sigma*\left(diag(\Sigma)\right)^{-1/2}$$ and therefore \autoref{matrix_VaR} can be used for VaR calculation.

In contrast to previous approaches, \textit{Portfolio UQR} is calculated in a different fashion. We firstly create portfolio returns and portfolio volatility series using individual returns and correlation structure obtained from Realized Covariance matrix, $\Sigma$, as $$r_{t,P}=w^{\top}*r_t $$ and $$\sigma_{t,P}=\sqrt{w^{\top}*\Sigma_t*w},$$ where $r_{t,P}$ and $\sigma_{t,P}$ is portfolio return and portfolio volatility at time $t$ respectively and $r_t$ is vector of individual returns at time $t$.   Series $r_{t,P}$ and $\sigma_{t,P}$ are further modeled using Univariate Quantile Regression Model for Returns and the forecasts of the quantiles of the portfolio return series are considered to be semi-parametric percentage portfolio VaR.   

\subsubsection{Statistical Evaluation}
In the statistical comparison we study absolute performance which tells us whether model is dynamically correctly specified, i.e. we study goodness-of-fit, and relative performance in which we compare models against each other. For the absolute performance evaluation we use modified version of the Dynamic Quantile test \citep{engle2004caviar}, referred to as the CAViaR test by \cite{berkowitz2011evaluating}. In their work, \cite{berkowitz2011evaluating} define ``hit'' variable in a way that 
$$ hit_{t+1}=
\left\{
\begin{array}{ll}
1 & if  \quad r_{t+1}\leq Q_{r_{t+1}}({\tau}) \\
0 & otherwise \\
\end{array}
\right.$$
i.e. $hit_{t+1}$ is a binary variable taking values 1 if conditional quantile is violated and 0 otherwise. Hit series of a dynamically correctly specified series should be i.i.d Bernoulli distributed with parameter $\tau$ $$ hit_{t+1}\sim iid(\tau,\tau(1-\tau))$$ By construction, $hit$ is a binary variable, therefore \cite{berkowitz2011evaluating} propose to test the hypothesis of correct dynamic specification using following logistic regression $$ hit_t=c+\sum_{d=1}^n\beta_{1d}hit_{t-l} + \sum_{d=1}^n\beta_{2d}Q_{r_{t-d+1}}(\tau)+u_t $$ where $u_t$ is assumed to have logistic distribution. We use likelihood ratio test to verify null hypothesis that $\beta$`s are equal to zero and $\mathbb{P}(hit_t=1)=\dfrac{e^c}{1+e^c}=\tau$. Exact finite sample critical values for the likelihood ratio test are obtained from Monte Carlo simulation as suggested by \cite{berkowitz2011evaluating}.

Relative performance of benchmark models is tested using expected tick loss for pairwise model comparison \citep{giacomini2005evaluation,clements2008quantile}. Loss function is defined as $$ \mathcal{L}_{\tau,m}= E \left(( \tau-I\left(e_{t+1}^{m}<0 \right)e_{t+1}^m\right),$$ where $I(\cdot)$ is indicator function, $e_{t+1}^{m}=r_{t+1}-Q_{r_{t+1}}^m(\tau)$ and $Q_{r_{t+1}}^m(\tau)$ is the $m$`th model quantile forecast. Forecasting accuracy of two models is  assessed using \cite{diebold2002comparing} test. Null hypothesis of the test that expected losses of two models are equal i.e. $H_0:\mathcal{L}_{\tau,1}=\mathcal{L}_{\tau,2}$ is tested against general alternative. 

\subsubsection{Economic Evaluation}
As we mentioned at the begining of the section besides the statistical evaluation we also study performance from the economic point of view which is particularly important for practitioners (e.g. portfolio managers). At first, for the economic evaluation of portfolio Value-at-Risk forecasts modified approach of \cite{Markowitz1952} is used. From the original work of \cite{Markowitz1952} it differs in a way that we concentrate on the relationship of the return and Value-at-Risk compared to original risk--return trade-off \footnote{Note that if we assume that quantiles of returns are standard normally distributed and we use standard cut--off points i.e. -1.645 for 5\% quantile both approaches are equivalent}. To overcome the difficulties of specifying proper model for returns and covariance/correlation matrices we decide to use their ex-post realizations i.e. for day $T$ we use returns realized in day $T$, realized covariance/correlation matrix in day $T$ and forecasts of univariate VaR for day $T$.  

In general, efficient frontier of the optimal portfolio can be constructed in a two equivalent ways:
\begin{enumerate}
\item Expected portfolio return is maximized for various levels of portfolio Value-at-Risk
\item Portfolio Value-at-Risk is minimized for various levels of expected portfolio return 
\end{enumerate}   

In both approaches asset weights, $w=(w_1,\hdots,w_q)'$, maximizing utility of risk averse investor can be found by solving following problem:  
\begin{eqnarray}
\underset{w_{t+1}}{\min} \quad w'_{t+1} \widehat{\Xi}_{t+1\vert t} w_{t+1}
\end{eqnarray}
$$ \mbox{s.t.} \quad l' w_{t+1}=1$$
$$ w'_{t+1} \geq 0 \footnote{We do not allow short-selling in this set-up.}$$
$$ w'_{t+1} \widehat{\mu}_{t+1}=\mu_P$$

where $w_{t+1}$ is $n\times 1$ vector of assets weights, $l$ denotes a $n\times 1$ vector of ones, $\widehat{\mu}_{t+1}$ is a vector of ex-post returns, $\mu_P$ stands for portfolio return and $\widehat{\Xi}_{t+1\vert t}=diag\left(\widehat{\%VaR_{t+1\vert t}}\right)* \widehat{\Omega}_{t+1}* diag\left(\widehat{\%VaR}_{t+1\vert t} \right)$ represents a correlated Value-at-Risk covariance matrix where $\widehat{\%VaR_{t+1\vert t}}$ is $n\times 1$ vector of univariate \%VaR forecast and $\widehat{\Omega}_{t+1}$ is correlation matrix obtained from realized covariance matrix estimate. Once we solve optimization problem for different levels of risk we construct efficient frontier. In the Markowitz-type portfolio optimization exercise we do not allow short-selling in order to meet restrictions imposed mainly by regulators on certain types of investors (pension funds etc.).

Finally we get to description of the second economic evaluation criteria used in our study, Global Minimum Value-at-Risk Portfolio. Basic problem of GMVaRP is similar to Markowitz, there are only two differences in the set-up. The first one is the existence of the closed-form solution. As a consequence we are not restricting asset weights because global minimum of the optimization problem might require negative weights of some assets. Second difference is the absence of targeted portfolio return. Therefore in some cases we might get negative portfolio return for the asset weights minimizing the overall risk of the portfolio. GMVaRP optimization problem can be written as   
\begin{eqnarray}
\underset{w_{t+1}}{\min} \quad w'_{t+1} \widehat{\Xi}_{t+1\vert t} w_{t+1}
\end{eqnarray}
$$ \mbox{s.t.} \quad l' w_{t+1}=1.$$
In the \cite{Kempf2006a} was shown that analytic solution of the problem is
\begin{eqnarray}
\label{eq:GMVaRP}
w_{t+1}^{GMVaR}=\frac{\widehat{\Xi}_{t+1\vert t}^{-1}l}{l' \widehat{\Xi}_{t+1\vert t}^{-1}l}, 
\end{eqnarray}
and portfolio Value-at-Risk corresponding to calculated asset weights is finally obtained as    
$$ {\%VaR}_{t+1}^{GMVaR}={w_{t+1}^{GMVaR}}'\widehat{\Xi}_{t+1\vert t}w_{t+1}^{GMVaR}.$$

\section{Simulation Study} \label{sim_study}
Before we analyze empirical data we would like to show performance of the newly proposed model in the controlled environment. Our aim is to show how various error distributions used for continuous price process simulation affect performance of the Panel Quantile Regression for Returns model.  

  As it is common in the literature let's assume that price process follow jump diffusion process with stochastic volatility:
\begin{equation}
\label{eq:sim_price}
\begin{split}
dp_t=\left(\mu -\frac{\sigma^2_t}{2}\right)dt +\sigma_tdW_{1t}+c_tdN_t  \\
d\sigma^2_t=\kappa\left(\alpha-\sigma^2_t\right)dt+\gamma\sigma_tdW_{2t},
\end{split}
\end{equation}
where $W_1$ and $W_2$ are Brownian motions, $c_tdN_t$ is a compound Poisson process with random jump size distributed as $N(0,\sigma_J)$ and $\sigma_J=0.01$. Parameters in \autoref{eq:sim_price} are set to the values which are reasonable for a stock price, i.e. $\alpha=0.04$, $\kappa=5$, $\gamma=0.5$ as in \cite{Zhang2005} and $\mu=0$ because we assume that returns are zero-mean. The volatility parameters satisfy Feller's condition $2\kappa\alpha\geq\gamma^2$, which keeps the volatility process away from the zero
boundary. Moreover we assume that $W_1$ comes from one of the following distributions with $\Sigma$ being Realized Covariance matrix obtained from the empirical data.
\begin{enumerate}
\item Multivariate normal distribution, $N(0,\Sigma)$.
\item Multivariate Student-t distribution with 9 degrees of freedom, $t_9(0,\Sigma)$.
\item Univariate normal distribution, $N(0,1)$.
\item Univariate Student-t distribution with 9 degrees of freedom, $t_9(0,1)$.
\end{enumerate}

To work with similar environment as with empirical data, we simulate 7 hours of 1 minutes intra-day prices for 2613 days. From the intra-day prices we calculate daily returns and all the realized measures. In case of multivariate normal and multivariate Student-t  distribution we use empirical estimate of Realized Covariance matrix for given day as the input for multivariate random number generation. For each error distribution we run 500 simulations. In each simulation step we use same estimation procedure as in case of empirical data - rolling window of length 1000.
 
\subsection{In-Sample Fit}
We start description of the results with the data generated from Multivariate Normal Distribution i.e. $N(0,\Sigma)$. For the rest of the distributions results are presented in the \nameref{appendix} - \autoref{tab:sim_UND_insample_PQR}, \ref{tab:sim_MTD_insample_PQR}, \ref{tab:sim_UTD_insample_PQR}  and we comment here only main differences from  Multivariate Normal Distribution. \autoref{tab:sim_MND_insample_PQR} shows detailed estimation results for 5\%, 10\%, 90\% and 95\% quantiles that are most important from the economic point of view for all three model specifications. To get a better view of quantile dynamics we also report lower and upper quartile together with median.

Table \ref{tab:sim_MND_insample_PQR} reveals significant estimates (except median) for PQR-RV model, with parameter values increasing in quantiles. Median coefficient is zero as a consequence of setting $\mu$ in \autoref{eq:sim_price}. Similarly to PQR-RV model, all but median quantiles are statistically significant also for the second model, PQR-RSV. We can notice difference in smaller magnitudes of coefficients in comparison to PQR-RV. Since both positive and negative semivariance should carry equal information in Multivariate Normal distribution, we expect equal coefficients. Finally, PQR-BPV model shows insignificant estimates for jump component, while coefficients for the volatility component are equal to PQR-RV model. This again is consistent with our expectation, as simulated jump variation in the simulations is too small. We conclude with observation that results for all three models are symmetric, as expected. 

Tables \ref{tab:sim_UND_insample_PQR}, \ref{tab:sim_MTD_insample_PQR}, and \ref{tab:sim_UTD_insample_PQR} reveal similar patterns. Heavy tails introduced to the data with Student-t distribution cause higher coefficients on both tails.

\begin{table}[h!]
\begin{center}
\small
\caption{Multivariate Normal Distribution -- Mean of coefficients estimates from Monte-Carlo simulations} \label{tab:sim_MND_insample_PQR}
\begin{tabular}{lccccccc}
\toprule
$\tau$ & 5\% & 10\% & 25\% & 50\% & 75\% & 90\% & 95\% \\ \cmidrule{2-8} \\[-0.75em]
 & \multicolumn{7}{c}{\textit{PQR-RV}} \\ \cmidrule{2-8}
$\hat{\beta}_{RV^{1/2}}$ & -1.54 & -1.15 & -0.57 & 0 & 0.56 & 1.14 & 1.55 \\ 
 & (-18.9) & (-18.37) & (-12.51) & (-0.06) & (11.85) & (17.72) & (17.97) \\  \cmidrule{2-8} \\[-0.75em]
 & \multicolumn{7}{c}{\textit{PQR-RSV}} \\ \cmidrule{2-8} 
$\hat{\beta}_{{RS^{+}}^{1/2}}$ & -1.12 & -0.83 & -0.43 & -0.02 & 0.36 & 0.76 & 1.04 \\ 
 & (-2.17) & (-2.35) & (-2.23) & (-0.18) & (1.96) & (2.29) & (2.07) \\[0.5em]
$\hat{\beta}_{{RS^{-}}^{1/2}}$ & -1.06 & -0.79 & -0.38 & 0.02 & 0.44 & 0.86 & 1.15 \\ 
 & (-2.06) & (-2.21) & (-1.95) & (0.15) & (2.42) & (2.63) & (2.32) \\ \cmidrule{2-8} \\[-0.75em]
 & \multicolumn{7}{c}{\textit{PQR-BPV}} \\ \cmidrule{2-8} 
$\hat{\beta}_{BPV^{1/2}}$ & -1.55 & -1.15 & -0.57 & 0 & 0.57 & 1.15 & 1.55 \\ 
 & (-18.9) & (-18.46) & (-12.49) & (-0.06) & (11.83) & (17.81) & (17.87) \\[0.5em]
$\hat{\beta}_{Jumps^{1/2}}$ & 0.06 & 0.04 & 0.02 & 0 & -0.03 & -0.05 & -0.06 \\ 
 & (0.49) & (0.56) & (0.45) & (0.01) & (-0.47) & (-0.63) & (-0.52) \\ 
\bottomrule
\end{tabular}
\begin{tablenotes}
\centering
\footnotesize
\item{Note: Table displays mean of coefficient estimates with corresponding t-statistics in parentheses. Individual fixed effects $\alpha_i(\tau)$ are not reported for brevity.}
\end{tablenotes}
\end{center}
\end{table}

\subsection{Out-of-Sample Performance}
In the out-of-sample forecasting exercise we start with comparison of absolute performance represented by various measures of unconditional coverage ($\widehat{\tau}_{avg}$, $\widehat{\tau}_{max}$, $\widehat{\tau}_{min}$, $\widehat{\tau}_{avg-dev}$) and dynamic quantile CAViaR test ($\widehat{DQ}_{violations}$) followed by pair-wise relative comparison according to Diebol-Mariano test ($DM$). For the unconditional coverage we report average unconditional coverage ($\widehat{\tau}_{avg}$) from the Monte-Carlo simulation which indicates how close our model was to theoretical quantile hit rate (i.e. for 5\% quantile we expect unconditional coverage to be somewhere around 5\%), maximum and minimum unconditional coverage ($\widehat{\tau}_{max}$, $\widehat{\tau}_{min}$) which show the range of possible movements of unconditional coverage rate and average deviation from the theoretical quantile hit rate ($\widehat{\tau}_{avg-dev}$) that shows on average how close our estimates were to theoretical values. Results of Diebold-Mariano test shows us percentage values when the benchmark model was outperformed by its competitors.

In the $Panel A.1$ and $Panel A.2$ of the \autoref{tab:sim_MND} we present absolute performance of the PQR and benchmark models respectively. Overall we can say that all the models are dynamically correctly specified in the majority of simulation trials for all the quantiles but median. Models with the lowest average deviation from studied quantile $\tau$ are all PQR specifications and UQR for all but median quantile. In case of median Portfolio UQR is the winner. Similarly to in-sample fit we obtain qualitatively identical results when we study data simulated from Multivariate Student-t distribution. When we switch to univariate error distributions situation change a bit and the Portfolio UQR seems to be the model with lowest average deviation and with lowest number of dynamically not correctly specified models. However, we must stress that for all but median quantile all the results are close to each other which indicates that none of the models is systematically misspecified.

More interesting comparison comes from $Panel B.1$ and $Panel B.2$ of the \autoref{tab:sim_MND} where PQR models are compared to benchmarks directly. All the PQR variants outperform significantly Portfolio UQR in all studied quantiles and RiskMetrics in all quantiles but median. Median RiskMetrics performance is overall the best which we attribute to the fact that median cut-off point for VaR calculation is zero and by construction series we simulated are supposed to be zero mean. When we concentrate on the comparison of the PQR to UQR situation is identical in both tails - UQR outperform slightly all PQR specifications. We address this result to the nature of simulated data - data generating process is driven by generated random numbers and contains just little heterogeneity that could possibly translate into the gains using PQR. Median performance however is better for PQR which is result of the averaging in the PQR median calculation. Moreover as the number of estimated parameters is significantly lower in case of PQR compared to UQR, median forecasts are less noisy which translates to better median PQR performance directly. Qualitatively identical results are obtained also for Multivariate Student-t distribution. If we turn to comparison with univariate distributions PQR outperform UQR significantly in all studied quantiles. Source of this interesting fact lies in the degree of heterogeneity present in the data. The only source of heterogeneity in our simulated data is random number generation process. In case of univariate distributions each generated time series has errors that are independent from remaining time series. However, in the multivariate distributions all error terms are affected by each other because we assume some correlation/covariance structure. As a result multivariate random numbers are more homogeneous compared to univariate one.

Generally results obtained from Monte-Carlo simulation helps us to justify the use of panel quantile regression for modelling quantiles of future returns. Our main results are that whatever error distribution for simulation we use PQR models are specified well dynamically, they dominate RiskMetrics and Portfolio UQR benchmark models and are slightly outperformed by UQR due to the lack of heterogeneity in the simulated data in case of multivariate distributions while PQR outperform UQR in more heterogeneous data created by univariate error distributions. We also show importance of covariance structure in the comparison of the results of multivariate and univariate distributions.    

\rotatebox{90}{\parbox[c][15cm]{\textheight}{%
\centering
\captionof{table}{Models performance using data simulated from Multivariate Normal Distribution}
\resizebox{1.3\textwidth}{!}{ 
    \begin{tabular}{lcccccccccccccccccc}
    \toprule
          &       & \multicolumn{5}{c}{PQR-RV}            &       & \multicolumn{5}{c}{PQR-RSV}           &       & \multicolumn{5}{c}{PQR-BPV} \\
\cmidrule{3-7}\cmidrule{9-13}\cmidrule{15-19}    \textit{Panel A.1} &       & 5\%   & 10\%  & 50\%  & 90\%  & 95\%  &       & 5\%   & 10\%  & 50\%  & 90\%  & 95\%  &       & 5\%   & 10\%  & 50\%  & 90\%  & 95\% \\
    \midrule
          & $\widehat{DQ}_{violations}$ & 6.2   & 4.8   & 12.2  & 2.8   & 7.8   &       & 6.8   & 5     & 13    & 2.8   & 7.8   &       & 6.2   & 5.2   & 12.4  & 2.6   & 7 \\
          & $\widehat{\tau}_{avg}$ & 5.0   & 10.1  & 51.4  & 90.1  & 95.1  &       & 5.0   & 10.1  & 51.4  & 90.1  & 95.0  &       & 5.1   & 10.1  & 51.4  & 90.1  & 95.0 \\
          & $\widehat{\tau}_{max}$ & 6.5   & 11.8  & 55.4  & 91.6  & 96.4  &       & 6.5   & 11.7  & 55.6  & 91.7  & 96.4  &       & 6.3   & 11.7  & 55.4  & 91.7  & 96.2 \\
          & $\widehat{\tau}_{min}$ & 3.6   & 8.4   & 47.5  & 88.5  & 93.7  &       & 3.5   & 8.3   & 47.4  & 88.5  & 93.7  &       & 3.5   & 8.5   & 47.4  & 88.5  & 93.6 \\
          & $\widehat{\tau}_{avg-dev}$ & 0.0   & 0.1   & 1.4   & 0.1   & 0.1   &       & 0.0   & 0.1   & 1.4   & 0.1   & 0.0   &       & 0.1   & 0.1   & 1.4   & 0.1   & 0.0 \\
          &       &       &       &       &       &       &       &       &       &       &       &       &       &       &       &       &       &  \\
          &       & \multicolumn{5}{c}{RiskMetrics}       &       & \multicolumn{5}{c}{UQR}               &       & \multicolumn{5}{c}{Portfolio UQR} \\
\cmidrule{3-7}\cmidrule{9-13}\cmidrule{15-19}    \textit{Panel A.2} &       & 5\%   & 10\%  & 50\%  & 90\%  & 95\%  &       & 5\%   & 10\%  & 50\%  & 90\%  & 95\%  &       & 5\%   & 10\%  & 50\%  & 90\%  & 95\% \\
    \midrule
          & $\widehat{DQ}_{violations}$ & 9.2   & 14    & 6.6   & 18.8  & 7.4   &       & 6.8   & 5.2   & 14.2  & 3     & 8.2   &       & 7     & 4.8   & 3     & 4.2   & 7 \\
          & $\widehat{\tau}_{avg}$ & 5.2   & 9.3   & 50.5  & 91.1  & 95.0  &       & 5.0   & 10.1  & 51.5  & 90.1  & 95.1  &       & 4.8   & 9.6   & 49.9  & 90.4  & 95.2 \\
          & $\widehat{\tau}_{max}$ & 7.1   & 11.2  & 54.1  & 92.7  & 96.4  &       & 6.3   & 11.7  & 55.6  & 91.8  & 96.4  &       & 6.1   & 11.5  & 52.5  & 91.8  & 96.3 \\
          & $\widehat{\tau}_{min}$ & 3.3   & 7.1   & 46.5  & 89.3  & 93.6  &       & 3.5   & 8.4   & 47.7  & 88.5  & 93.5  &       & 3.7   & 8.1   & 47.4  & 88.5  & 94.0 \\
          & $\widehat{\tau}_{avg-dev}$ & 0.2   & -0.7  & 0.5   & 1.1   & 0.0   &       & 0.0   & 0.1   & 1.5   & 0.1   & 0.1   &       & -0.2  & -0.4  & -0.1  & 0.4   & 0.2 \\
          &       &       &       &       &       &       &       &       &       &       &       &       &       &       &       &       &       &  \\
          &       & \multicolumn{17}{c}{benchmark} \\
\cmidrule{3-19}          &       & \multicolumn{5}{c}{RiskMetrics}       &       & \multicolumn{5}{c}{UQR}               &       & \multicolumn{5}{c}{Portfolio UQR} \\
\cmidrule{3-7}\cmidrule{9-13}\cmidrule{15-19}    \textit{Panel B.1} &       & 5\%   & 10\%  & 50\%  & 90\%  & 95\%  &       & 5\%   & 10\%  & 50\%  & 90\%  & 95\%  &       & 5\%   & 10\%  & 50\%  & 90\%  & 95\% \\
    \midrule
    PQR-RV & $DM$  & 58.8  & 62.6  & 0.2   & 65.6  & 54.8  &       & 3.4   & 2.6   & 9.4   & 4.4   & 4.2   &       & 47.4  & 43.6  & 20.4  & 44.8  & 48.6 \\
    PQR-RSV & $DM$  & 58.4  & 62.2  & 0.2   & 64.4  & 54    &       & 2.8   & 2     & 9.4   & 2.6   & 3     &       & 46.6  & 42.6  & 20.4  & 44.2  & 47.2 \\
    PQR-BPV & $DM$  & 58    & 61.8  & 0.2   & 63.8  & 54    &       & 1.6   & 1     & 8.8   & 1.6   & 1.2   &       & 44.2  & 40.4  & 20.6  & 42.4  & 44.8 \\
          &       &       &       &       &       &       &       &       &       &       &       &       &       &       &       &       &       &  \\
          &       & \multicolumn{5}{c}{PQR-RV}            &       & \multicolumn{5}{c}{PQR-RSV}           &       & \multicolumn{5}{c}{PQR-BPV} \\
\cmidrule{3-7}\cmidrule{9-13}\cmidrule{15-19}    \textit{Panel B.2} &       & 5\%   & 10\%  & 50\%  & 90\%  & 95\%  &       & 5\%   & 10\%  & 50\%  & 90\%  & 95\%  &       & 5\%   & 10\%  & 50\%  & 90\%  & 95\% \\
    \midrule
    RiskMetrics & $DM$  & 0.2   & 0     & 13.8  & 0     & 0     &       & 0.2   & 0     & 14.2  & 0     & 0     &       & 0.2   & 0     & 13    & 0     & 0 \\
    UQR   & $DM$  & 6.8   & 8.8   & 2.6   & 9.2   & 9     &       & 12    & 13.6  & 2.8   & 15.2  & 12.2  &       & 8.8   & 11.2  & 3     & 12.4  & 12 \\
    Portfolio UQR & $DM$  & 0.2   & 1     & 0.6   & 0     & 0.6   &       & 0.2   & 1.6   & 0.6   & 0     & 0.6   &       & 0.2   & 1     & 0.6   & 0     & 0.8 \\
    \bottomrule
    \end{tabular}%
    }
\begin{tablenotes}
\footnotesize
\item{Note: Table displays absolute and relative performance of PQR models for returns with RV, RSV and BPV as regressors and benchmark models. All values are in \%.}
\medskip
\item {\textit{Panel A.1} reports absolute performance of PQR models, \textit{Panel A.2} reports absolute performance of  benchmark models. For each model and quantile $\tau$, percentage of violations of the CAViaR test for correct dynamic specification ($\widehat{DQ}_{violations}$), average unconditional coverage ($\widehat{\tau}_{avg}$),maximum unconditional coverage ($\widehat{\tau}_{max}$), minimum unconditional coverage ($\widehat{\tau}_{min}$) and average deviation of unconditional coverage from given quantile $\tau$ ($\widehat{\tau}_{avg-dev}$)}
\medskip
\item{\textit{Panel B.1} and \textit{Panel B.2} report relative performance of Panel Quantile Regression Models for Returns in comparison to benchmark models and relative performance of benchmark models in comparison to Panel Quantile Regression Models for Returns respectively. For each specification and quantile $\tau$ we report percentage of statistically better performance according to Diebold-Mariano($DM$) test at 5\% significance level.}
\end{tablenotes}
  \label{tab:sim_MND}%
}}

\section{Empirical Application}\label{empir app}

Confident about performance of the modeling strategy in the controlled environment, we turn to application of the proposed models on the empirical data. First, we describe in-sample fit of the PQR-RV, PQR-RSV and PQR-BPV model specifications. Second, we present results of the out-of-sample Value--at--Risk forecasting exercise. Third, we complement results of statistical evaluation by the simple portfolio allocation exercise where we study Global Minimum Value-at-Risk Portfolio and Markowitz like relationship between Value-at-Risk and return of the portfolio.

Empirical application is carried out using 29 U.S. stocks\footnote{Apple Inc. (AAPL), Amazon.com, Inc. (AMZN), Bank of America Corp (BAC), Comcast Corporation (CMCSA), Cisco Systems, Inc. (CSCO), Chevron Corporation (CVX), Citigroup Inc. (C), Walt Disney Co (DIS), General Electric Company (GE), Home Depot Inc. (HD), International Business Machines Corp. (IBM), Intel Corporation (INTC), Johnson \& Johnson (JNJ), JPMorgan Chase \& Co. (JPM), The Coca-Cola Co (KO), McDonald's Corporation (MCD), Merck \& Co., Inc. (MRK),Microsoft Corporation (MSFT), Oracle Corporation (ORCL), PepsiCo, Inc. (PEP), Pfizer Inc. (PFE), Procter \& Gamble Co (PG), QUALCOMM, Inc. (QCOM), Schlumberger Limited. (SLB), AT\&T Inc. (T), Verizon Communications Inc. (VZ), Wells Fargo \& Co (WFC), Wal-Mart Stores, Inc. (WMT), Exxon Mobil Corporation (XOM).} that are traded at New York Stock Exchange. These stocks have been chosen according to market capitalization and their liquidity.  Sample we study spans from July 1, 2005 to December 31, 2015 and we consider trades executed within U.S. business hours (9:30 -- 16:00 EST). In order to ensure sufficient liquidity and eliminate possible bias we explicitly exclude weekends and bank holidays (Christmas, New Year's Day, Thanksgiving Day, Independence Day). In total, our final dataset consists of 2613 trading days. Basic descriptive statistic of the data can be found in \autoref{tab:descriptive_stat} in \nameref{appendix}.

For estimation and forecasting purposes we use rolling window estimation with fixed length of 1000 observation,\footnote{We have tried different length of rolling window with the qualitative results of our analysis remaining unchanged. These results are available from authors upon request.} hence our model is always calibrated on a 4 years history. Our analysis is restricted to 5 minutes intraday log-returns that are used for computation of the daily returns and realized measures.   

\begin{figure}[H]

\centering
\caption{PQR-RV parameter estimates: $\lambda=1$} 
\includegraphics[scale=0.5]{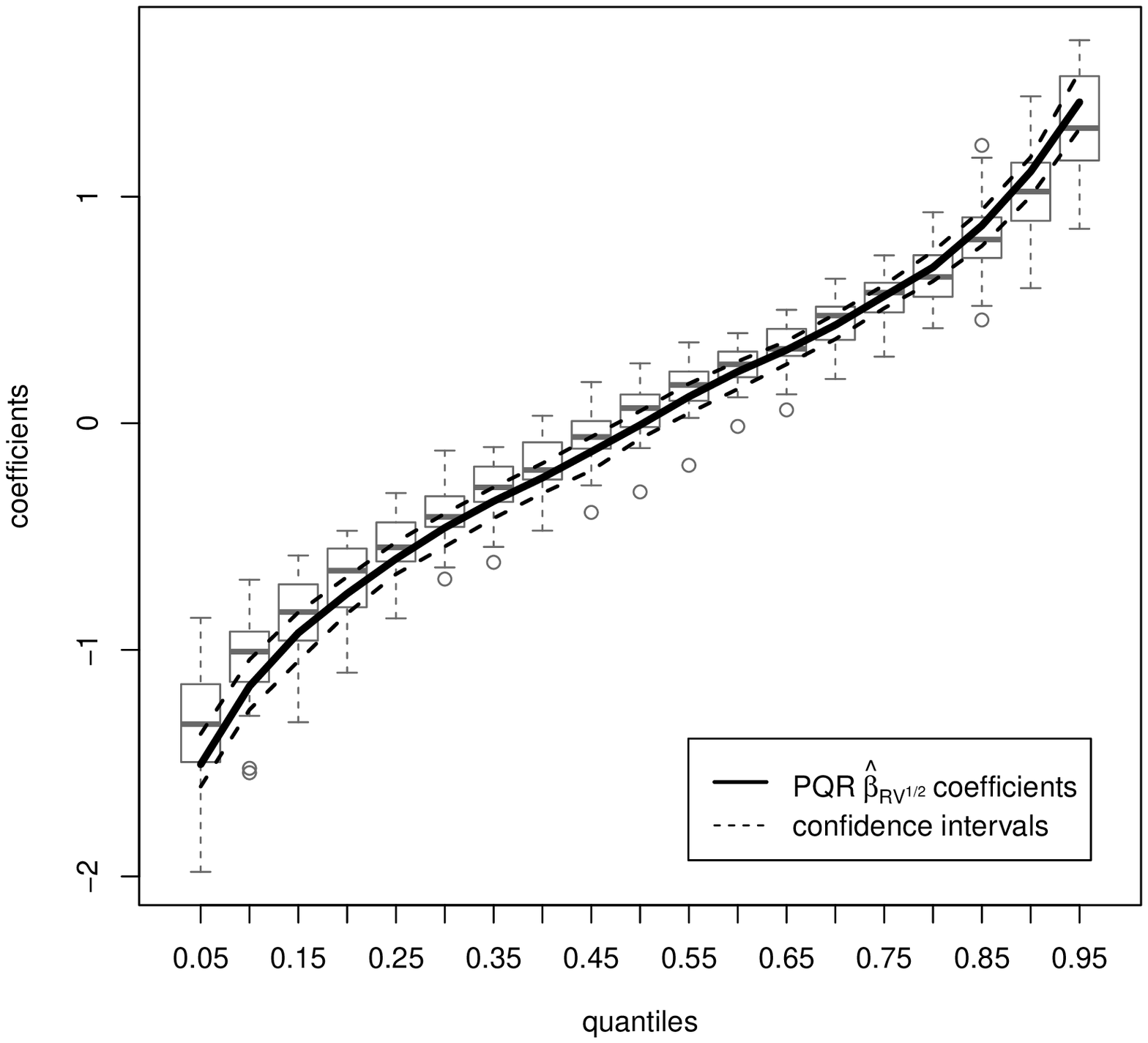}\label{fig:PQR-RV_lambda_1}
\captionsetup{justification=centering}
\floatfoot{Note: Parameter estimates with corresponding 95\% confidence intervals from the PQR-RV specification are ploted by solid and dashed lines respectively. Individual UQR-RV estimates are ploted in boxplots.} 
\end{figure}

\begin{figure}[H]
\centering
\caption{PQR-RSV parameter estimates: $\lambda=1$} 
\begin{subfigure}[b]{0.425\textwidth}
\caption{${RS^+}^{1/2}$}
\includegraphics[scale=0.4]{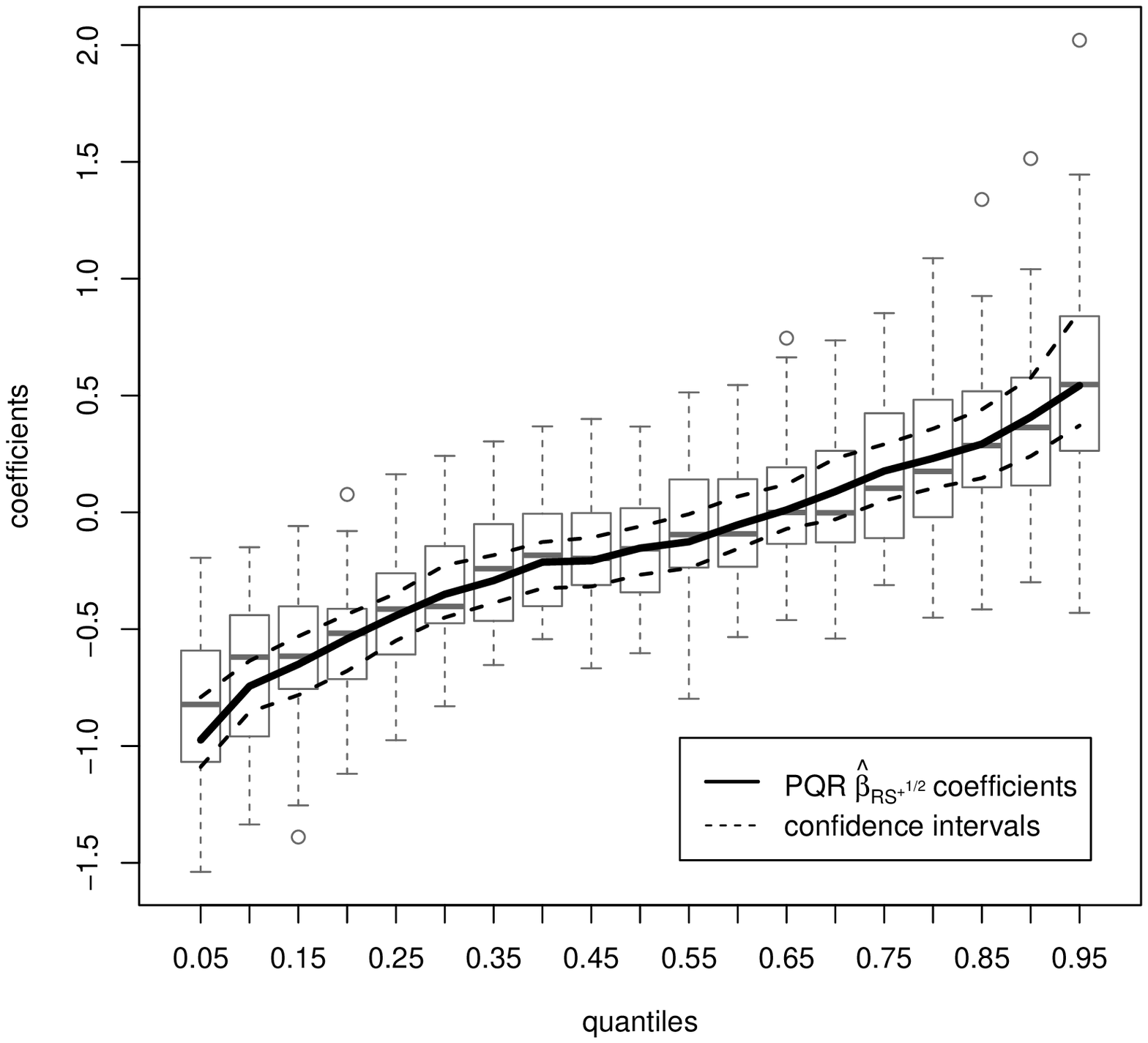}
\end{subfigure}
\begin{subfigure}[b]{0.425\textwidth}
\caption{${RS^-}^{1/2}$}
\includegraphics[scale=0.4]{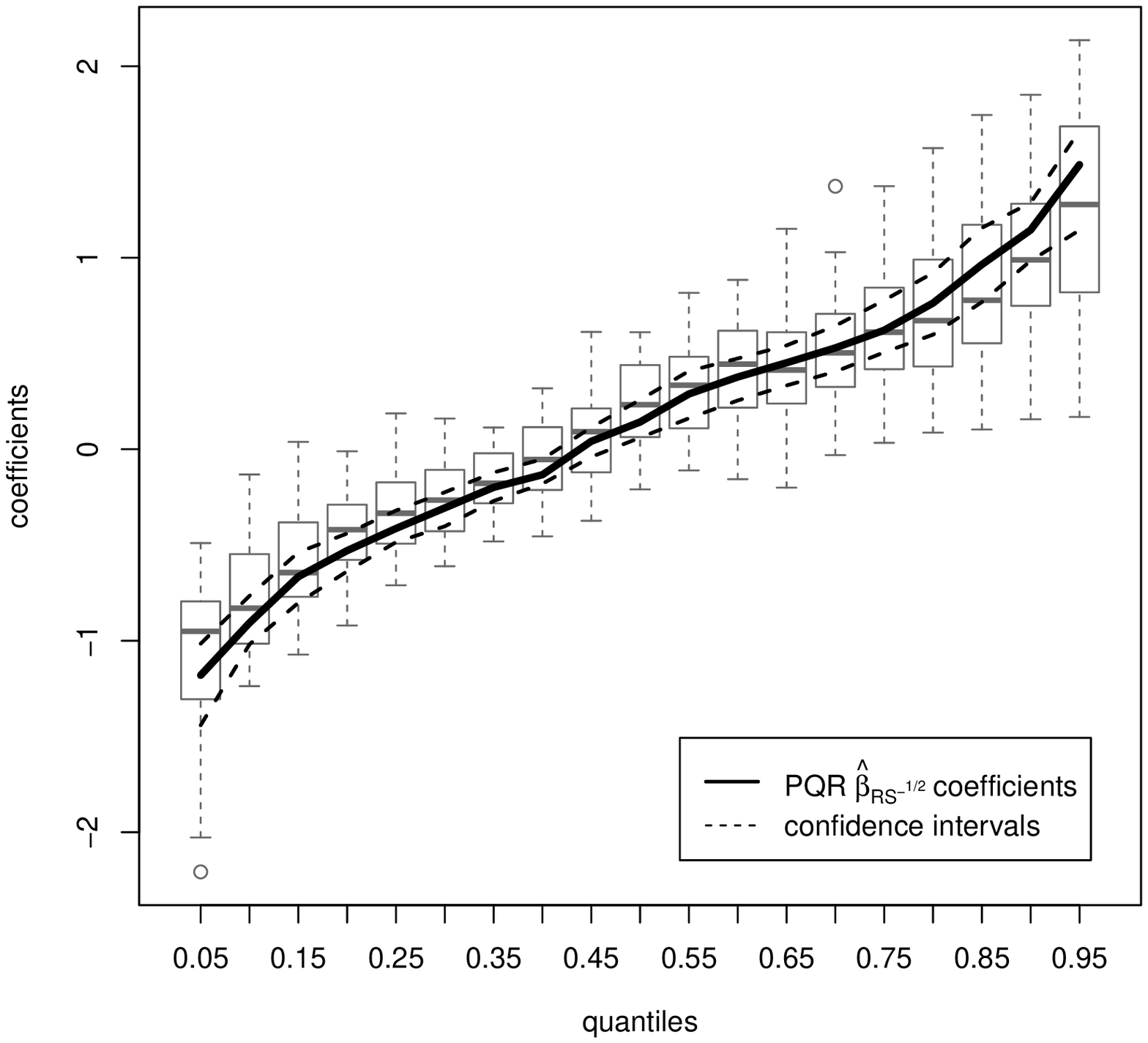}
\end{subfigure}
\captionsetup{justification=centering}
\floatfoot{Note: For both realized upside and downside semivariance parameters estimates with corresponding 95\% confidence intervals are ploted by solid and dashed lines respectively. Individual UQR-RSV estimates are ploted in boxplots.}
\label{fig:PQR-RSV_lambda_1}
\end{figure}

\begin{figure}[H]
\centering
\caption{PQR-BPV parameter estimates: $\lambda=1$} 
\begin{subfigure}[b]{0.425\textwidth}
\caption{${BPV}^{1/2}$}
\includegraphics[scale=0.4]{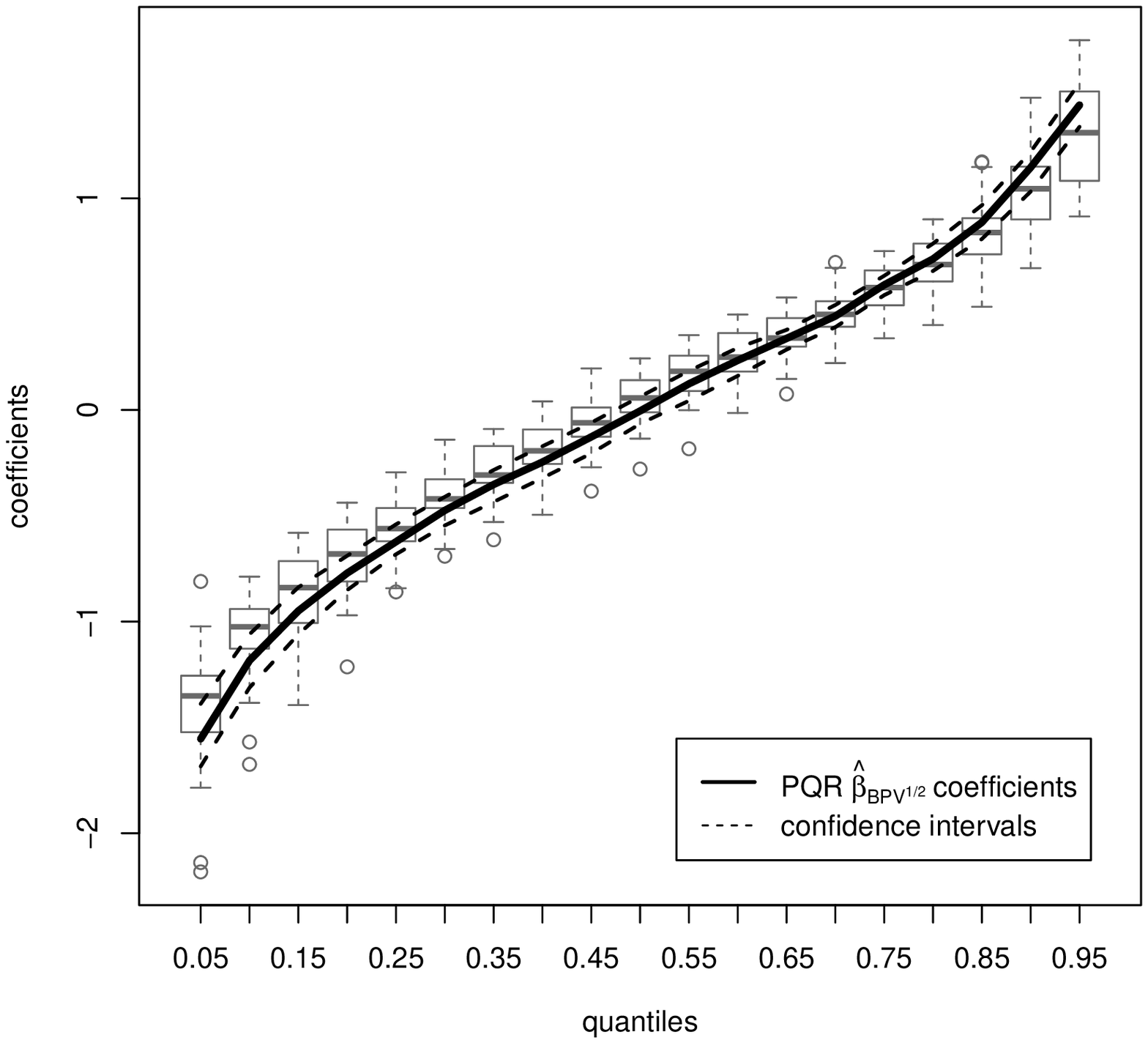}
\end{subfigure}
\begin{subfigure}[b]{0.425\textwidth}
\caption{${Jumps}^{1/2}$}
\includegraphics[scale=0.4]{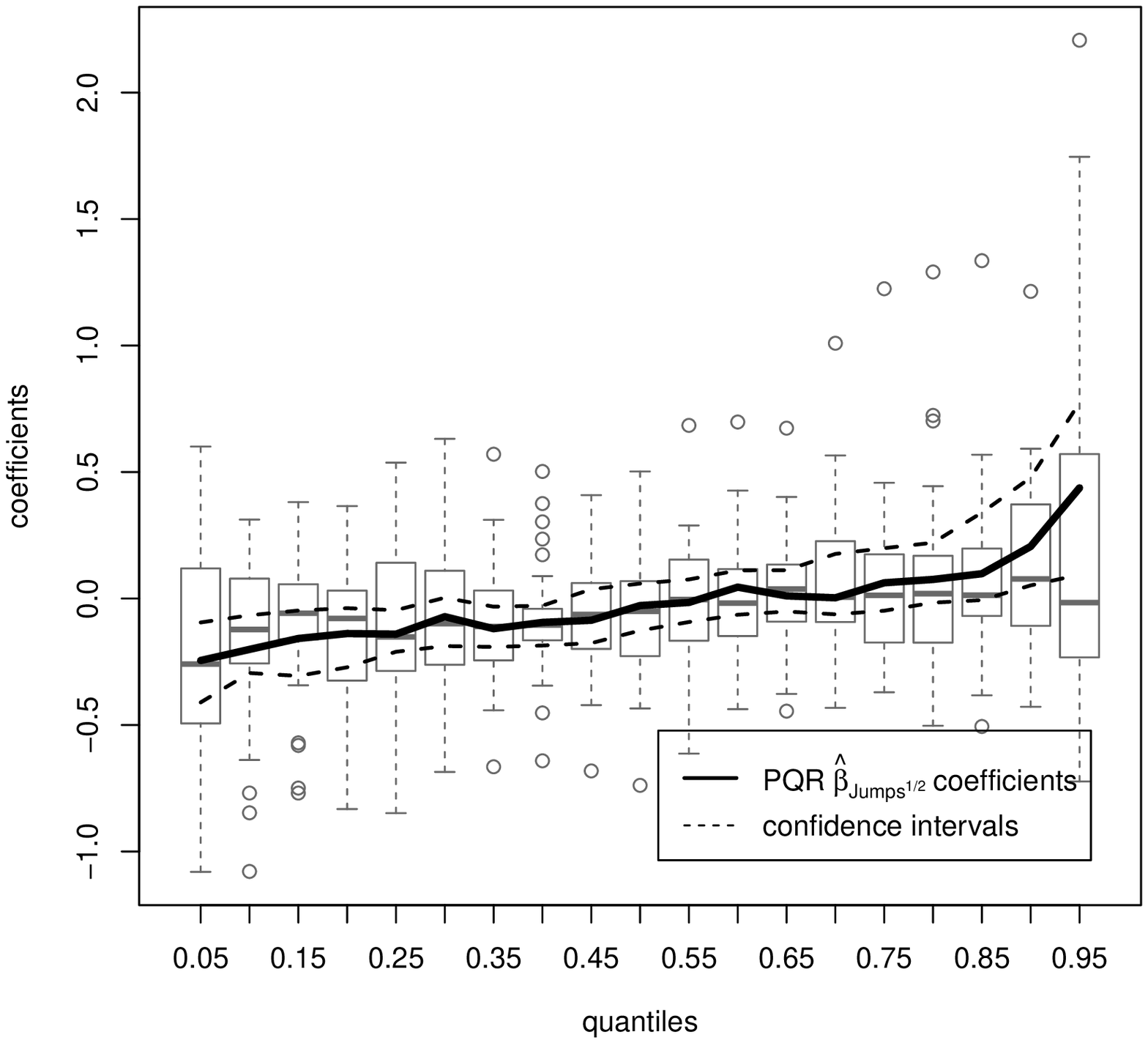}
\end{subfigure}
\captionsetup{justification=centering}
\floatfoot{Note: For both realized bi-power variation and jump component parameters estimates with corresponding 95\% confidence intervals are ploted by solid and dashed lines respectively. Individual UQR-BPV estimates are ploted in boxplots}
\label{fig:PQR-BPV complete lambda 1} 
\end{figure}

All the results presented in this section were obtained using pure fixed effects panel quantile regression, i.e. penalty parameter $\lambda$ set to 0. In the \nameref{appendix}\footnote{\autoref{tab:insample_PQR_lambda_1}, \autoref{fig:PQR-RV_lambda_1}, \autoref{fig:PQR-RSV_lambda_1} and \autoref{fig:PQR-BPV complete lambda 1}} we present also estimation results when $\lambda=1$ which serves as a robustness check.
 
\subsection{In-Sample Fit}
\begin{figure}[H]
\centering
\caption{PQR-RV parameter estimates} 
\includegraphics[scale=0.5]{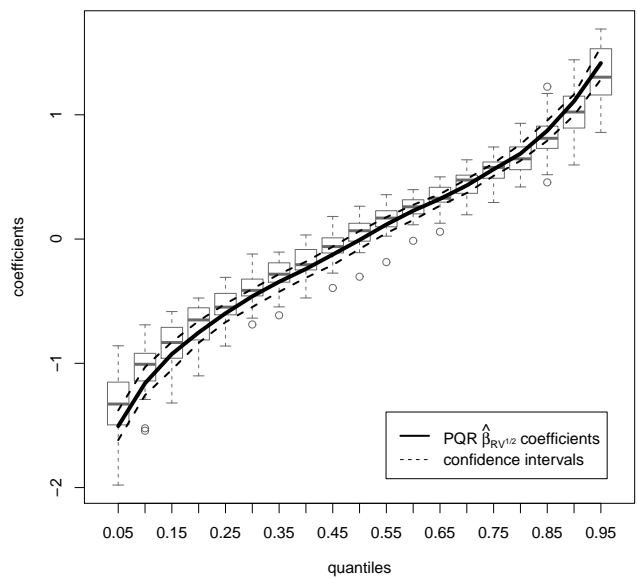}
\captionsetup{justification=centering}
\floatfoot{Note: Parameter estimates with corresponding 95\% confidence intervals from the PQR-RV specification are ploted by solid and dashed lines respectively. Individual UQR-RV estimates are ploted in boxplots.} \label{fig:PQR-RV}
\end{figure}

\begin{figure}[H]
\centering
\caption{PQR-RSV parameter estimates} 
\begin{subfigure}[b]{0.425\textwidth}
\caption{${RS^+}^{1/2}$}\label{fig:PQR-RS+}
\includegraphics[scale=0.4]{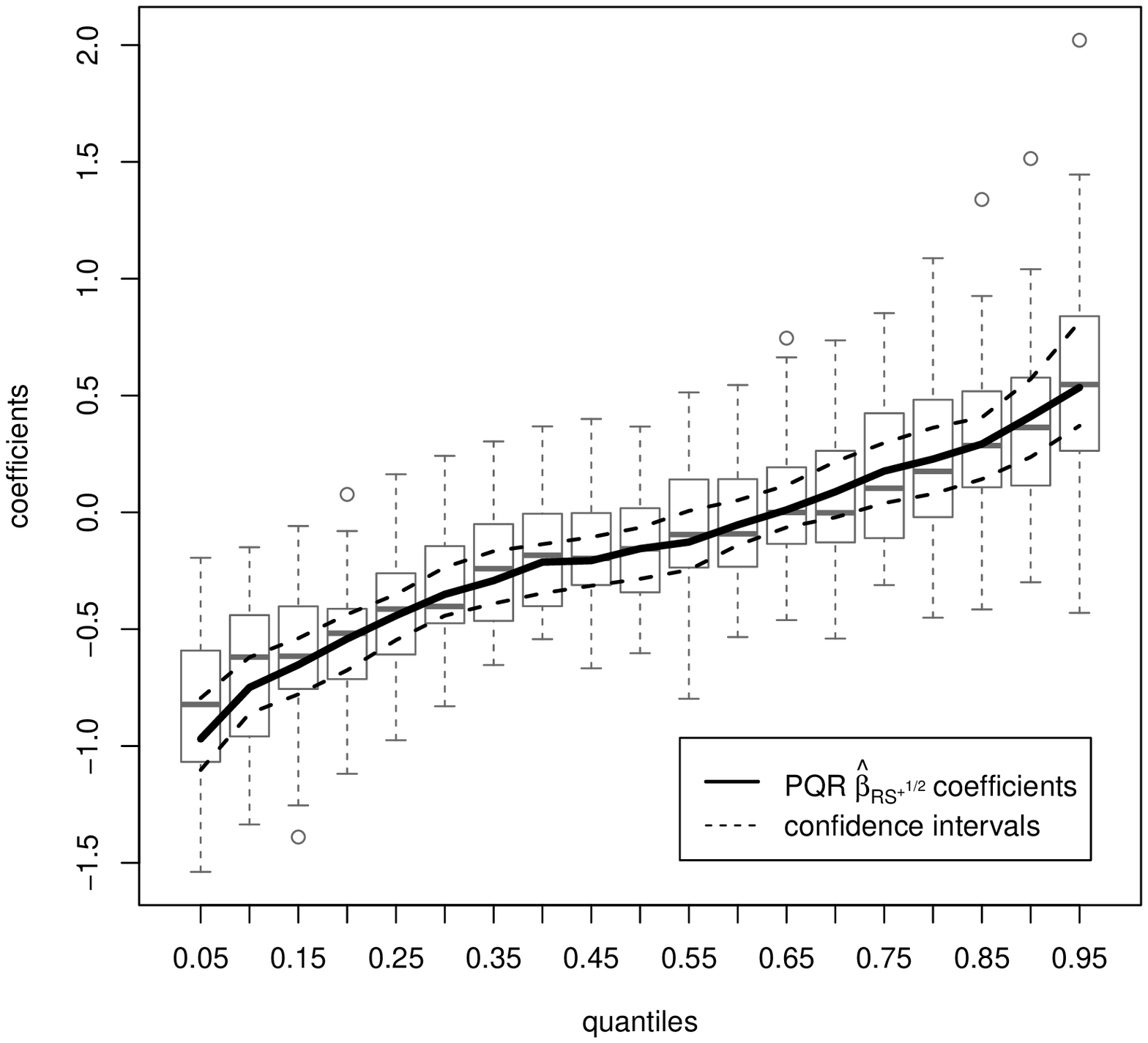}
\end{subfigure}
\begin{subfigure}[b]{0.425\textwidth}
\caption{${RS^-}^{1/2}$}\label{fig:PQR-RS-}
\includegraphics[scale=0.4]{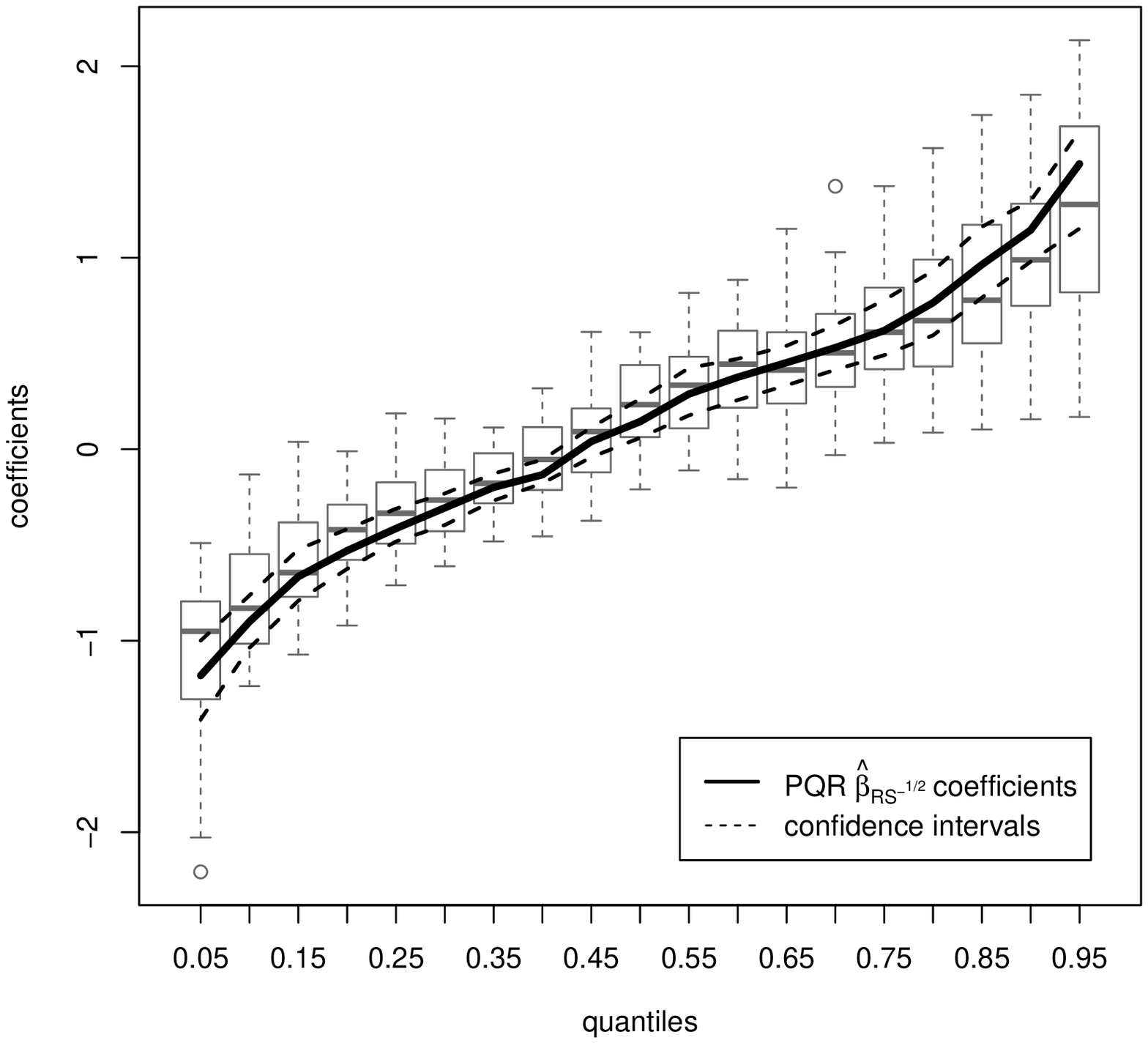}
\end{subfigure}
\captionsetup{justification=centering}
\floatfoot{Note: For both realized upside and downside semivariance parameters estimates with corresponding 95\% confidence intervals are ploted by solid and dashed lines respectively. Individual UQR-RSV estimates are ploted in boxplots.}
\label{fig:PQR-RSV}
\end{figure}

\begin{figure}[H]
\centering
\caption{PQR-BPV parameter estimates} 
\begin{subfigure}[b]{0.425\textwidth}
\caption{${BPV}^{1/2}$}\label{fig:PQR-BPV}
\includegraphics[scale=0.4]{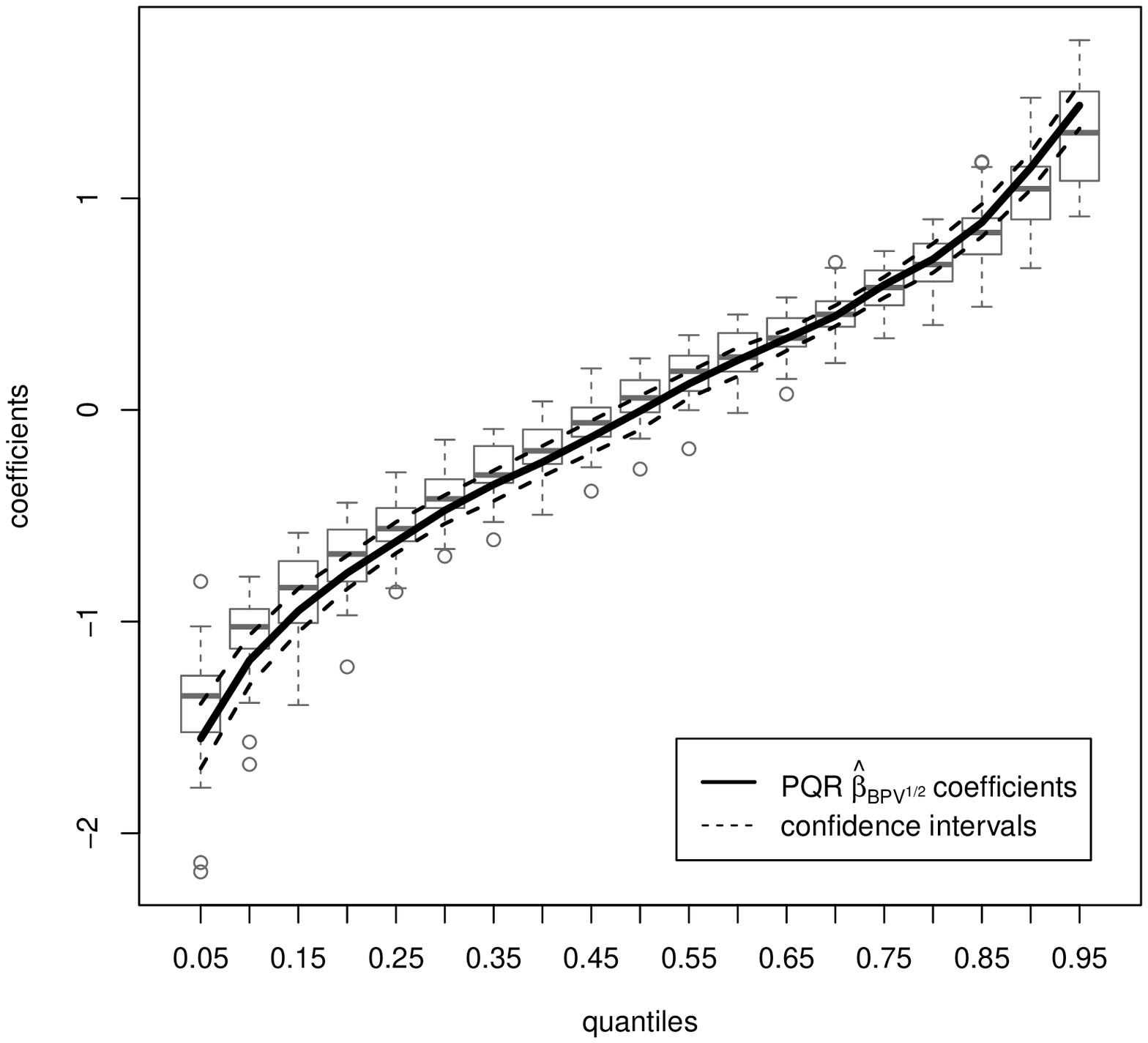}
\end{subfigure}
\begin{subfigure}[b]{0.425\textwidth}
\caption{${Jumps}^{1/2}$}\label{fig:PQR-BPV Jumps}
\includegraphics[scale=0.4]{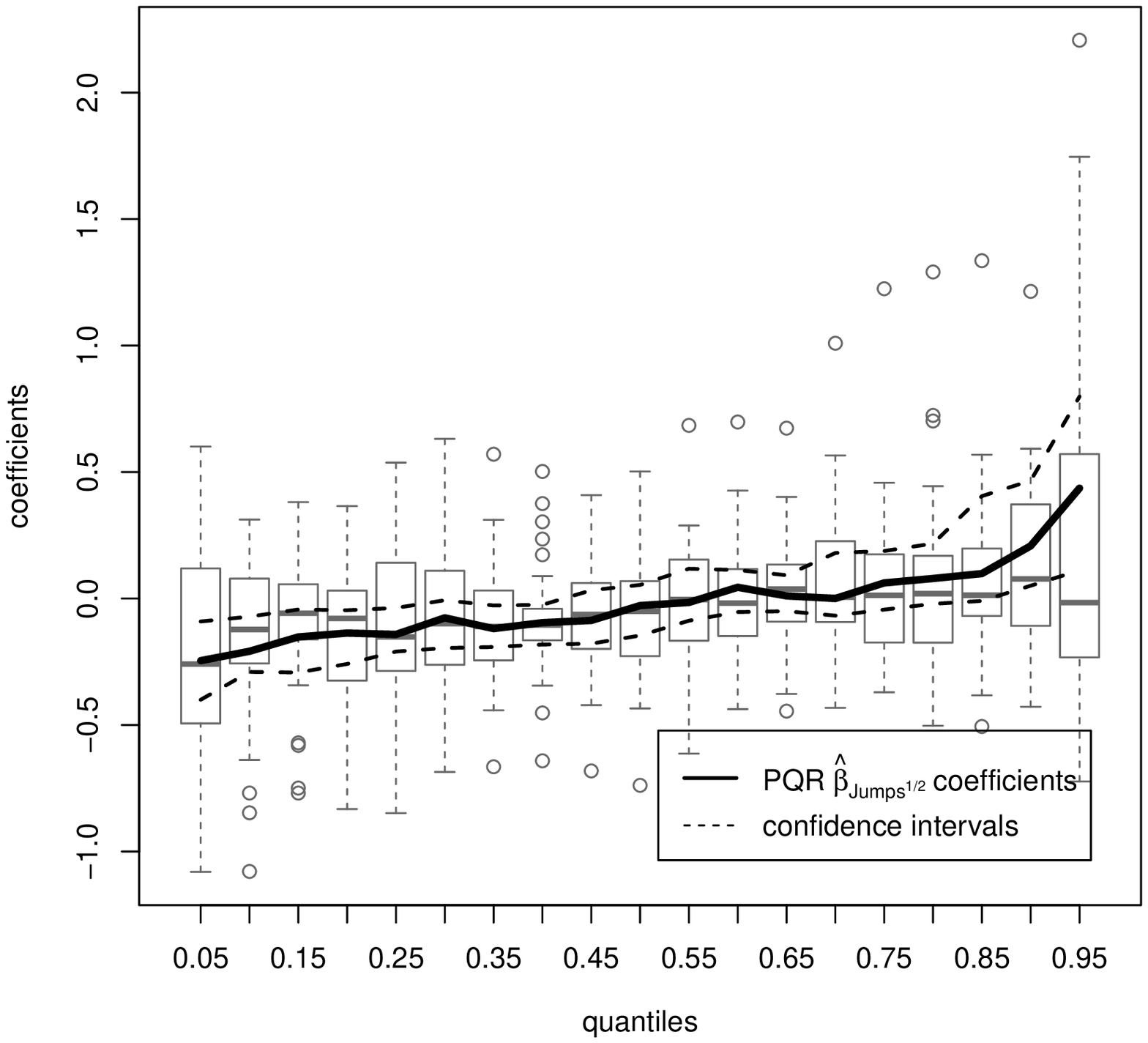}
\end{subfigure}
\captionsetup{justification=centering}
\floatfoot{Note: For both realized bi-power variation and jump component parameters estimates with corresponding 95\% confidence intervals are ploted by solid and dashed lines respectively. Individual UQR-BPV estimates are ploted in boxplots} \label{fig:PQR-BPV complete}
\end{figure}

Estimation results are detailed in the \autoref{tab:insample_PQR}. In addition, to get a better view of the dynamics, we show results of the PQR-RV, PQR-RSV and PQR-BPV also graphically in the Figures \ref{fig:PQR-RV}, \ref{fig:PQR-RSV} and \ref{fig:PQR-BPV complete} respectively. 

\begin{table}[H]
\begin{center}
\small
\caption{Coefficient estimates of Panel Quantile Regressions} \label{tab:insample_PQR}
\begin{tabular}{lccccccc}
\toprule
$\tau$ & 5\% & 10\% & 25\% & 50\% & 75\% & 90\% & 95\% \\ \cmidrule{2-8} \\[-0.75em]
 & \multicolumn{7}{c}{\textit{PQR-RV}} \\ \cmidrule{2-8}
$\hat{\beta}_{RV^{1/2}}$ & -1.5 & -1.16 & -0.6 & -0.01 & 0.56 & 1.11 & 1.42 \\ 
 & (-23.5) & (-20.62) & (-15.65) & (-0.2) & (20.37) & (24.84) & (20.7) \\  \cmidrule{2-8} \\[-0.75em]
 & \multicolumn{7}{c}{\textit{PQR-RSV}} \\ \cmidrule{2-8} 
$\hat{\beta}_{{RS^{+}}^{1/2}}$   &-0.97 & -0.75 & -0.44 & -0.16 & 0.18 & 0.41 & 0.53 \\ 
 & (-12.74) & (-11.98) & (-8.31) & (-2.73) & (2.69) & (4.55) & (4.51) \\[0.5em]
$\hat{\beta}_{{RS^{-}}^{1/2}}$ & -1.18 & -0.9 & -0.41 & 0.14 & 0.62 & 1.14 & 1.49 \\ 
 & (-11.72) & (-14.05) & (-9.9) & (2.7) & (9.17) & (13.66) & (10.39) \\ \cmidrule{2-8} \\[-0.75em]
 & \multicolumn{7}{c}{\textit{PQR-BPV}} \\ \cmidrule{2-8} 
$\hat{\beta}_{BPV^{1/2}}$ & -1.55 & -1.18 & -0.62 & 0 & 0.59 & 1.15 & 1.44 \\ 
 & (-19.5) & (-18.15) & (-16.27) & (-0.13) & (23.84) & (23.22) & (25.72) \\[0.5em]
$\hat{\beta}_{Jumps^{1/2}}$ & -0.25 & -0.21 & -0.14 & -0.03 & 0.06 & 0.21 & 0.44 \\ 
 & (-3.24) & (-3.54) & (-3.39) & (-0.58) & (1.11) & (1.9) & (2.56) \\ 
\bottomrule
\end{tabular}
\begin{tablenotes}
\centering
\footnotesize
\item{Note: Table displays coefficient estimates with bootstraped t-statistics in parentheses. Individual fixed effects $\alpha_i(\tau)$ are not reported for brevity - they are available from authors upon request}
\end{tablenotes}
\end{center}
\end{table}

\autoref{tab:insample_PQR} reveals that parameters of the first model specification (PQR-RV) where lagged volatility is used to explain conditional quantiles of returns are significantly different from zero for all quantiles except median. Moreover, signs of the estimated parameters correspond to our expectations -- coefficients at lower (upper) quantiles are negative (positive). Our expectations follow the Value--at--Risk concept in which quantiles of standard normal distribution are combined with volatility estimate. For illustration, 5\% and 95\% quantiles of standard normal distribution are -1.645 and 1.645 respectively. Furthermore, median parameter estimate that is not statistically significant confirm hypothesis about the randomness/unpredictability of the short-term returns. 

In the \autoref{tab:insample_PQR}, we can also see that absolute values of parameter estimates are not symmetric around median which highlight the relative importance of the realized volatility on the estimation of the lower quantiles of returns. We arrive to a similar conclusion also when looking at the \autoref{fig:PQR-RV} that compares and displays PQR estimates together with their corresponding 95\% confidence intervals and individual UQR parameter estimates plotted in boxplots. Importantly, the \autoref{fig:PQR-RV} shows that once we control for unobserved heterogeneity by PQR past volatility has larger influence on both the lower and the upper quantiles of returns than the majority of individual UQR. This is highlighted in far quantiles, e.g. coefficient of PQR in 5\% quantile is -1.5 whereas median of individual UQR coefficient is -1.33 (mean -1.36) or 95\% quantile PQR coefficient is 1.42 and median of individual UQR is only 1.30 (mean 1.31).   

This finding constitutes important empirical result, as we document unobserved heterogeneity in far quantiles that needs to be controlled. 

Coefficients form the second model specification (PQR-RSV) where Realized Variance is decomposed into realized downside ($RS^-$) and upside ($RS^+$) semivariance are significantly different from zero for all considered quantiles. Magnitude of coefficients driving impact of both variables is highest at far quantiles showing strongest impact of both negative, and positive semivariance on tails of the returns distributions. 

However, influence of $RS^-$ is far more important in the upper quantiles where it dominates $RS^+$. On the contrary, in the lower quantiles values of parameters are close to each other and therefore we cannot draw the similar conclusion as in upper quantiles. Median performance is bit different from PQR-RV case. We can see that coefficients for both $RS^-$ and $RS^+$ are statistically significant and in case that magnitude of $RS^-$ and $RS^+$ is equal they sum to -0.02 which translates into loss in 50\% of cases. However as theory and stylized facts about financial time series suggest influence of negative returns and subsequently negative semivariances should be greater than the effect of the positive one. Therefore one can not draw straightforward conclusion about sign and magnitude of median return. 

Careful reader might also notice that median coefficient of $\hat{\beta}_{{RS^{+}}^{1/2}}$ is negative and opposite is true for $\hat{\beta}_{{RS^{-}}^{1/2}}$. Explanation of this feature rely on short and long term mean-reversion nature of the returns and fact that we are using lagged values of realized semivariances as regressors. If we put it together negative return at day $t-1$ will cause that $RS^-_{t-1}>RS^+_{t-1}$ and prediction of the median quantile for day $t$ will be positive because $\hat{\beta}_{{RS^{-}}^{1/2}}$ is positive and vice versa for positive return and subsequent $RS^-_{t-1}<RS^+_{t-1}$. Behavior described in the previous sentence lead to mean-reversion. Results of our analysis are also supported by the \autoref{fig:PQR-RSV}. Similarly to PQR-RV specification we can see in the \autoref{fig:PQR-RSV} that controlling for unobserved heterogeneity among financial assets is important because influence of both downside and upside semivariance is greater in the lower quantiles than in individual UQR. For example in 5\% quantile coefficients obtained by PQR-RSV are -0.97 and -1.18 for $RS^+$ and $RS^-$ respectively, however median values of individual UQR are -0.82 (mean -0.84) for $RS^+$ and -0.95 (mean -1.1) for $RS^-$. Moreover, in the upper quantiles of negative semivariance (\autoref{fig:PQR-RS-})PQR coefficients differs substantially from individual UQR (95\% quantile $\hat{\beta}_{{RS^{-}}^{1/2}}$ coefficient of 1.49 vs. individual UQR median/mean coefficient of 1.28/1.27), however, the opposite is true for $RS^+$ (95\% quantile $\hat{\beta}_{{RS^{+}}^{1/2}}$ coefficient of 0.54 vs. individual UQR median/mean coefficient of 0.55/0.55). These findings support previous conclusion that $RS^-$ influences future upper quantiles of returns more than $RS^+$.

Finally, \autoref{tab:insample_PQR} reveals interesting results about parameter estimates of the third model specification (PQR-BPV), where the Bi-Power Variation and Jump Component as regressors are used to drive the return quantiles. We can infer that jumps have significant impact on both far upper and lower quantiles of future returns. To be precise, magnitude of the jump coefficient $\hat{\beta}_{Jumps^{1/2}}$ is highest for 95\% quantile with the value of 0.44. For the remaining above median quantiles jumps are not statistically significant and therefore influence of Quadratic Variation reduces to Integrated Variance represented by Bi-Power Variation. We can observe opposite situation for the below median quantiles where $\hat{\beta}_{Jumps^{1/2}}$ coefficients are always significant. \autoref{fig:PQR-BPV complete} helps us to confirm results of our previous analysis also graphically. If we compare \autoref{fig:PQR-BPV} to \autoref{fig:PQR-RV} we get almost identical picture. Moreover in the \autoref{fig:PQR-BPV Jumps} we can see that from 45\% to 85\% quantiles confidence intervals of the jump component are getting wider and include zero. Once we combine these two findings we can state that for these quantiles Quadratic Variation reduces to Integrated Variance. In contrast none of the confidence intervals of the 5\% to 40\% quantiles contain zero which highlights the relative importance of the jump component in the  modelling lower future quantiles of returns.

Overall, results of the in-sample analysis show asymmetric impact of the regressors on the quantiles of future returns. This impact is higher in the below median quantiles. We have also found evidence for positive/negative news asymmetry. This asymmetry is the highest in the 95\% quantile (0.53 coefficient of $RS^+$ vs. 1.49 of $RS^-$) while 5\% quantile shows only little asymmetry (-0.97 in case of $RS^+$ vs. -1.18 for $RS^-$). In addition we show importance of jumps for below median and far above median quantiles. Importantly, we document unobserved heterogeneity in far quantiles. We have also tested all three models (PQR-RV, PQR-RSV, PQR-BPV) for correct dynamic specification and we have found that none of them is systematically misspecified.

\subsection{Out-of-Sample Performance}
After discussion results of the in-sample analysis we now turn to description of the out-of-sample forecasting exercises. Similarly to simulation study we are analyzing Value--at--Risk performance of the equally weighted portfolio. Results of our analysis are presented in the following way: firstly we shortly comment on the absolute performance of the PQR models, secondly absolute performance of the benchmark models is discussed and lastly we concentrate on the most interesting relative performance comparison of the PQR models with respect to the benchmark models. All the results are summarized in the Table \ref{tab:abs_rel_performance}.

Unconditional coverage $\widehat{\tau}$ shown in \textit{Panel A.1} and \textit{Panel A.2} of the Table \ref{tab:abs_rel_performance} reveals that almost all models underestimate risk as values of unconditional coverage are higher than corresponding quantiles $\tau$, with few exceptions. Median quantiles as well as 5\% quantile of the Portfolio UQR and 90\% quantile of the PQR-RSV  overestimate risk although unconditional coverage of 89.9\% might be better described as perfect fit for PQR-RSV. We must also stress here that deviation from nominal quantile rates is generally lower than 1\% and we can not reject hypothesis of correct unconditional coverage. 

If we turn to median performance we  can see that all the models overestimate the risk. Moreover we can see that deviations from the nominal quantiles are higher compared to off-median quantiles. We address this finding to the nature of financial time series especially to stylized fact about the unpredictability of the returns. More important, this result corresponds to our motivation of explaining quantiles of the cross-section of market returns instead of expected value. Further more this is in line with our previous results presented in the in-sample section where median estimates where not statistically significant.

If we concentrate on the correct dynamic specification of the models represented by CAViaR test in the second and third line of the \textit{Panel A.1} and \textit{A.2} we see that all the models in all quantiles are dynamically correctly specified except median of RiskMetrics. In this case we strongly reject null hypothesis of proper dynamic specification  given p-value$<$0.01. We attribute poor median RiskMetrics performance to the construction of \autoref{standard_VaR} where cut-off point at 50\% quantile, $\gamma_{50\%}$, is 0\footnote{Median of standard normal distribution is 0.}.

Relative performance of the PQR models is summarized in the \textit{Panel B}\footnote{For brevity we report in Table \ref{tab:abs_rel_performance} only pair-wise comparison against benchmark models, full matrix of pairwise comparison is available from authors upon request.}. Results of our analysis indicate good relative performance of PQR models. All three Panel Quantile model specifications (PQR-RV, PQR-RSV and PQR-BPV) significantly outperform RiskMetrics in all studied quantiles. Moreover, all PQR specifications consistently outperformed Portfolio UQR in upper quantiles and UQR in several quantiles i.e. PQR-RV outperform individual UQR estimates in 10\% quantile, however performance of PQR-RSV is significantly better in 95\% quantile and PQR-BPV delivers significantly more accurate forecasts than individual UQR in 5\% and 10\% quantiles. If we concentrate on the full pair-wise comparison, the most important is the performance of the UQR as the main competitor of the PQR specifications. In all of the studied quantiles UQR is not able to outperform any of the PQR specification. This fact highlights the importance to control for unobserved heterogeneity among the assets. If we move from comparison of PQR and UQR models interesting is the relative performance of the Portfolio UQR which outperform RiskMetrics only at 5\% and 10\% quantiles. In contrast, UQR similarly to PQR outperform RiskMetrics in all studied quantiles. These results reveal the importance of the asset specific contribution to overall future portfolio risk as approach of firstly aggregating data and secondly modeling them is not able to capture dynamics creating variation in distribution of future portfolio returns.

\begin{sidewaystable}[H]
\centering
\captionof{table}{Out-of-sample performance of various specifications of Panel Quantile Regression Model for Returns}
\label{tab:abs_rel_performance}
\resizebox{\textwidth}{!}{%
    \begin{tabular}{rcccccccccccccccccc}
    \toprule
          &       & \multicolumn{5}{c}{PQR-RV}            &       & \multicolumn{5}{c}{PQR-RSV}           &       & \multicolumn{5}{c}{PQR-BPV} \\
\cmidrule{3-7}\cmidrule{9-13}\cmidrule{15-19}    \multicolumn{1}{l}{\textit{Panel A.1}} &  $\tau$     & 5\%   & 10\%  & 50\%  & 90\%  & 95\%  &       & 5\%   & 10\%  & 50\%  & 90\%  & 95\%  &       & 5\%   & 10\%  & 50\%  & 90\%  & 95\% \\
    \midrule
          &$\widehat{\tau}$& 0.060 & 0.108 & 0.465 & 0.901 & 0.959 &       & 0.059 & 0.107 & 0.465 & 0.899 & 0.960 &       & 0.058 & 0.107 & 0.465 & 0.902 & 0.960 \\
          & $\widehat{DQ}$    & 8.917 & 3.373 & 10.157 & 6.939 & 5.686 &       & 8.180 & 3.339 & 10.129 & 1.476 & 9.152 &       & 7.956 & 4.625 & 10.157 & 5.298 & 6.210 \\
          & \textit{p-val} & 0.178 & 0.761 & 0.118 & 0.326 & 0.459 &       & 0.225 & 0.765 & 0.119 & 0.961 & 0.165 &       & 0.241 & 0.593 & 0.118 & 0.506 & 0.400 \\
          &       &       &       &       &       &       &       &       &       &       &       &       &       &       &       &       &       &  \\
          &       & \multicolumn{5}{c}{RiskMetrics}       &       & \multicolumn{5}{c}{UQR} &       & \multicolumn{5}{c}{Portfolio UQR} \\
\cmidrule{3-7}\cmidrule{9-13}\cmidrule{15-19}    \multicolumn{1}{l}{\textit{Panel A.2}} &  $\tau$     & 5\%   & 10\%  & 50\%  & 90\%  & 95\%  &       & 5\%   & 10\%  & 50\%  & 90\%  & 95\%  &       & 5\%   & 10\%  & 50\%  & 90\%  & 95\% \\
    \midrule
          & $\widehat{\tau}$ & 0.061 & 0.094 & 0.451 & 0.919 & 0.958 &       & 0.061 & 0.107 & 0.467 & 0.902 & 0.960 &       & 0.043 & 0.099 & 0.491 & 0.909 & 0.955 \\
          & $\widehat{DQ}$    & 9.652 & 3.096 & \underline{20.600} & 9.452 & 10.899 &       & 8.323 & 3.041 & 9.067 & 7.174 & 6.796 &       & 9.426 & 5.988 & 3.273 & 4.507 & 3.238 \\
          & \textit{p-val} & 0.140 & 0.797 & 0.002 & 0.150 & 0.092 &       & 0.215 & 0.804 & 0.170 & 0.305 & 0.340 &       & 0.151 & 0.425 & 0.774 & 0.608 & 0.778 \\
          &       &       &       &       &       &       &       &       &       &       &       &       &       &       &       &       &       &  \\
          &       & \multicolumn{17}{c}{benchmark} \\ \cmidrule{3-19}
          &       & \multicolumn{5}{c}{RiskMetrics}       &       & \multicolumn{5}{c}{UQR} &       & \multicolumn{5}{c}{Portfolio UQR} \\
\cmidrule{3-7}\cmidrule{9-13}\cmidrule{15-19}    \multicolumn{1}{l}{\textit{Panel B}} &  $\tau$     & 5\%   & 10\%  & 50\%  & 90\%  & 95\%  &       & 5\%   & 10\%  & 50\%  & 90\%  & 95\%  &       & 5\%   & 10\%  & 50\%  & 90\%  & 95\% \\
\midrule
    \multicolumn{1}{l}{PQR-RV} & DM    & \textbf{-2.430} & \textbf{-2.259} & \textbf{-3.347} & \textbf{-2.127} & \textbf{-1.935} &       & 0.125 & \textbf{-1.734} & 1.350 & 0.801 & -0.362 &       & -0.733 & -1.590 & -0.310 & \textbf{-2.053} & \textbf{-2.260} \\
          & p-val & 0.008 & 0.012 & 0.000 & 0.017 & 0.027 &       & 0.550 & 0.041 & 0.911 & 0.788 & 0.359 &       & 0.232 & 0.056 & 0.378 & 0.020 & 0.012 \\
          &       &       &       &       &       &       &       &       &       &       &       &       &       &       &       &       &       &  \\
    \multicolumn{1}{l}{PQR-RSV} & DM    & \textbf{-2.368} & \textbf{-2.249} & \textbf{-3.561} & \textbf{-2.367} & \textbf{-2.242} &       & 1.244 & -1.569 & -0.438 & -1.268 & \textbf{-2.023} &       & -0.558 & -1.580 & -0.758 & \textbf{-2.921} & \textbf{-3.099} \\
          & p-val & 0.009 & 0.012 & 0.000 & 0.009 & 0.012 &       & 0.893 & 0.058 & 0.331 & 0.102 & 0.022 &       & 0.289 & 0.057 & 0.224 & 0.002 & 0.001 \\
          &       &       &       &       &       &       &       &       &       &       &       &       &       &       &       &       &       &  \\
    \multicolumn{1}{l}{PQR-BPV} & DM    & \textbf{-2.540} & \textbf{-2.422} & \textbf{-3.304} & \textbf{-2.055} & \textbf{-1.851} &       & \textbf{-1.796} & \textbf{-1.887} & 1.424 & 0.839 & 0.703 &       & -1.191 & \textbf{-1.887} & -0.294 & \textbf{-1.978} & \textbf{-1.705} \\
          & p-val & 0.006 & 0.008 & 0.000 & 0.020 & 0.032 &       & 0.036 & 0.030 & 0.923 & 0.799 & 0.759 &       & 0.117 & 0.030 & 0.384 & 0.024 & 0.044 \\
    \bottomrule
    \end{tabular}
}
\begin{tablenotes}
\footnotesize
\item{Note: Table displays absolute and relative performance of PQR models for returns with RV, RSV and BPV as regressors and benchmark models.}
\medskip
\item {\textit{Panel A.1} reports absolute performance of PQR models, \textit{Panel A.2} reports absolute performance of  benchmark models. For each model and quantile $\tau$, unconditional coverage ($\widehat{\tau}$), the value of the CAViaR test for correct dynamic specification ($\widehat{DQ}$) with corresponding Monte Carlo based p-value and the value of loss function($\widehat{L}$)  is displayed. Not correctly dynamically specified models are underlined. }
\medskip
\item{\textit{Panel B} reports relative performance of Panel Quantile Regression Models for Returns. For each specification and quantile $\tau$ we report Diebold-Mariano test statistics for pairwise comparison with benchmark models($\widehat{DM}$) with corresponding p-value. Significantly more accurate forecasts with respect to benchmark models at the 5\% significance level are in bold. Full matrix of pairwise comparison is available from authors upon request}
\end{tablenotes}
\end{sidewaystable}

\subsection{Economic Evaluation}
In the last section of the empirical data analysis we would like to see if statistical gains also translate to economic value. We concentrate on the comparison of 3 models -- PQR-RV, UQR and RiskMetrics, and refrain from presenting results for PQR-RSV and PQR-BPV for brevity. The construction of the Portfolio UQR rule out economic evaluation in our set-up because asset weights will be set before applying quantile regression and therefore results will be driven by covariance structure only. 

We start description of the results by Global Minimum Value-at-Risk Portfolio followed by Markowitz like optimization where we show Value-at-Risk -- Return relationship. In both approaches we use annualized portfolio returns and annualized portfolio Value-at-Risks. In the GMVaRP comparison we focus on both left and right tail together with median because we do not set any constraints regarding asset weights - according to \autoref{eq:GMVaRP} GMVaRP has closed form solution. On the contrary Markowitz like optimization is purely numeric and does not offer closed form solution therefore we restrict our analysis on long only positions. As a result we concentrate on the left tail of the return distribution only which shows us potential loss of the investor.   

Results of the GMVaRP analysis are displayed in \autoref{tab:GMVaRP}. For all studied quantiles but median model with the best performance is PQR-RV followed by UQR. RiskMetrics ended last and we must note that for median quantile we were not able to calculate value of GMVaRP due to problem with singularity of the correlated Value-at-Risk matrix\footnote{If we set cut-off point in \autoref{standard_VaR} to zero we get singular matrix of zeros that is not invertible.}

\begin{table}[H]
  \centering
  \caption{Global Minimum Value-at-Risk Portfolio}
    \begin{tabular}{lccccc}
    \toprule
    $\tau$ & 5\%   & 10\%  & 50\%  & 90\%  & 95\% \\
    \midrule
    PQR-RV & \textbf{11.76} & \textbf{8.69} & 0.02  & \textbf{9.46}  & \textbf{12.37} \\
    UQR & 11.85 & 8.79 & \textbf{0.01}  & 9.52  & 12.43 \\
    RiskMetrics & 12.77 & 9.95 & NaN   & 9.95  & 12.77 \\
    \bottomrule
    \end{tabular}%
    \begin{tablenotes}
\centering
\footnotesize
\item{Note: Table displays absolute percentage values of Global Minimum Value-at-Risk Portfolio for given quantile $\tau$. Best model for given quantiles is reported in bold.}
\end{tablenotes}
  \label{tab:GMVaRP}%
\end{table}%

\begin{figure}[h!]
\centering
\caption{Value-at-Risk -- Return efficient frontiers} 
\begin{subfigure}[b]{0.49\textwidth}
\caption{$5\% VaR$} 
\includegraphics[scale=0.6]{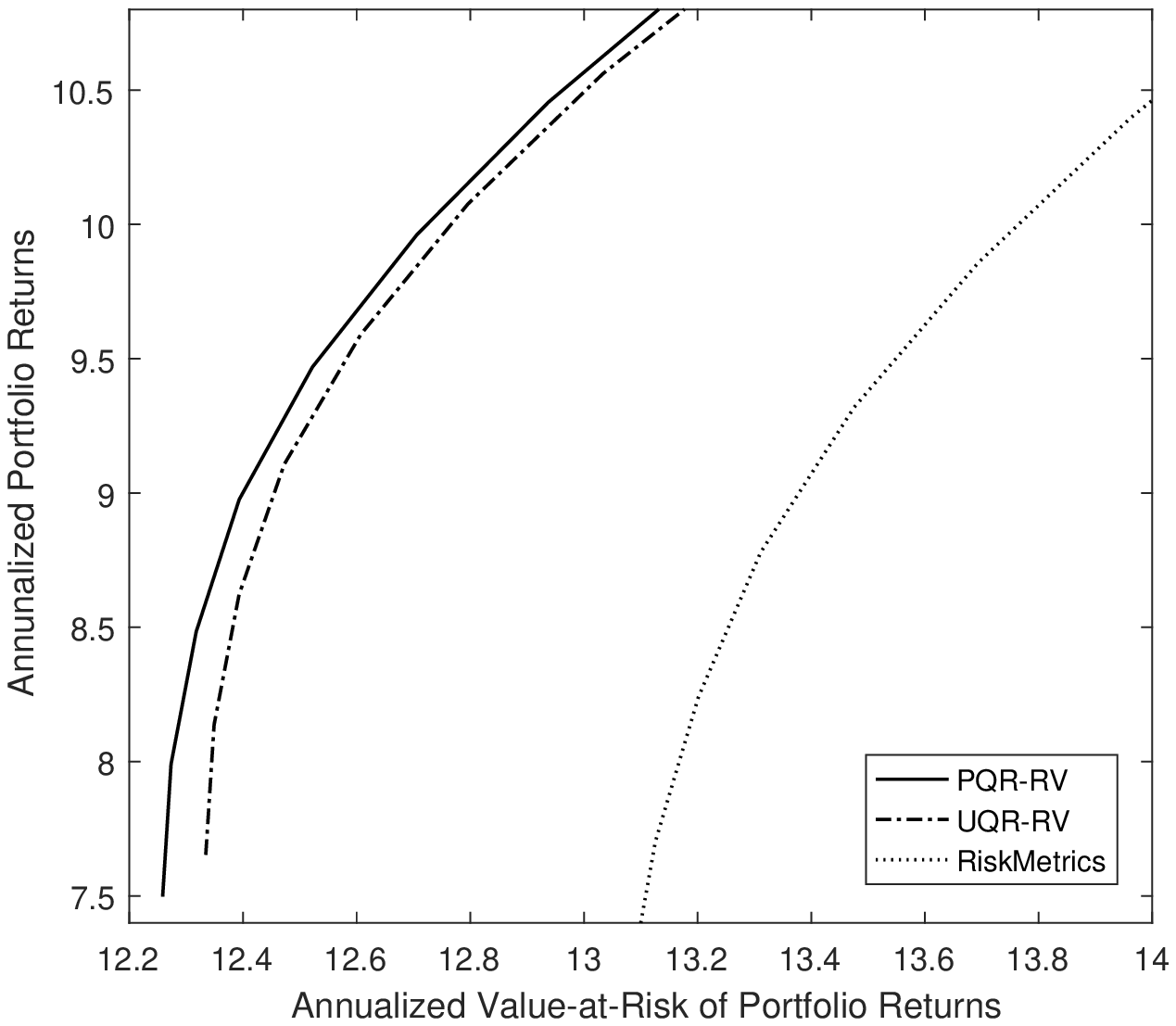}\label{fig:VaR_5}
\end{subfigure}
\begin{subfigure}[b]{0.49\textwidth}
\caption{$10\% VaR$}
\includegraphics[scale=0.6]{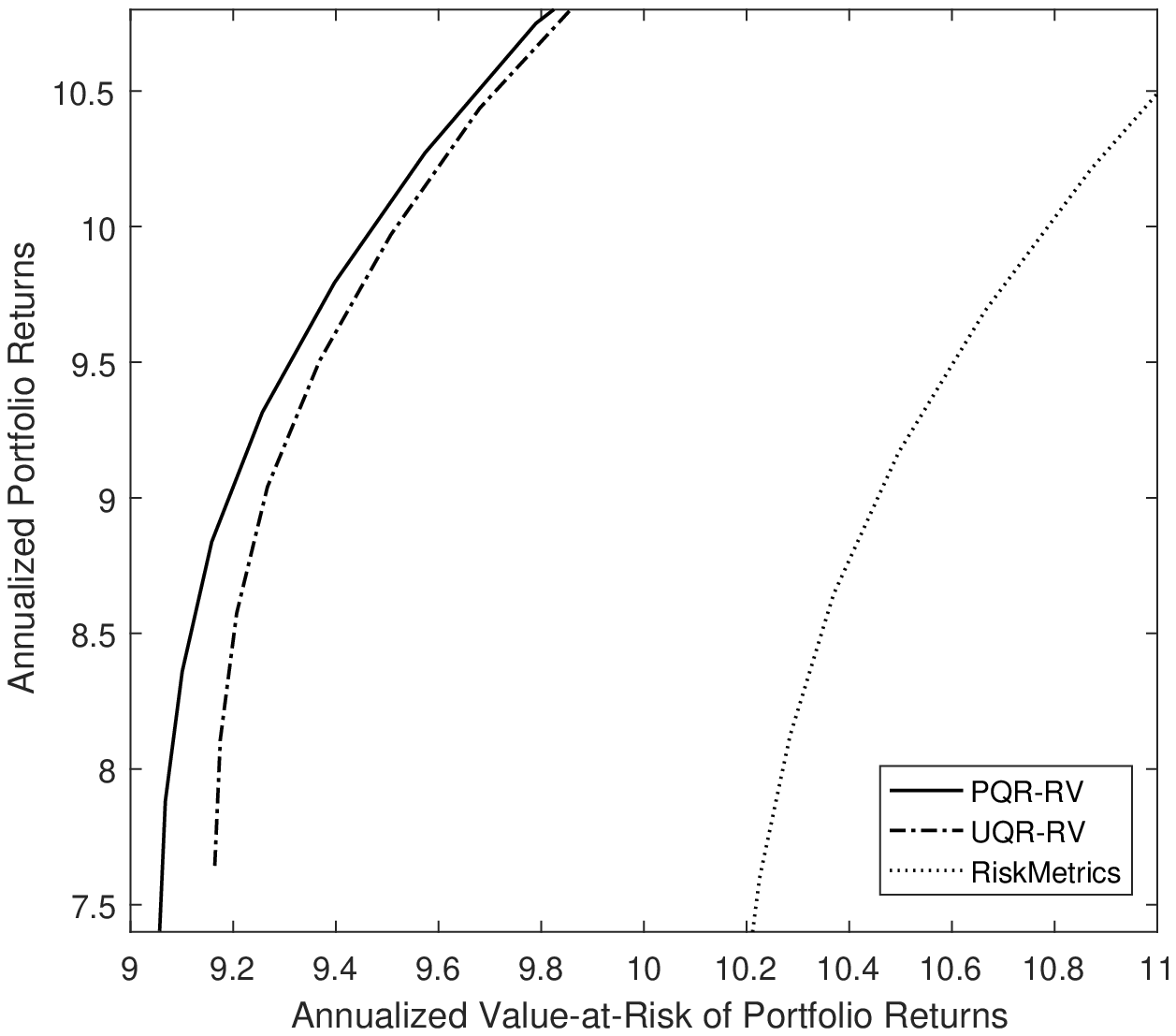}\label{fig:VaR_10}
\end{subfigure}
\captionsetup{justification=centering}
\floatfoot{Note: Percentage values of portfolio VaR and returns are displayed.}
\label{fig:VaR-ret eff}
\end{figure}

Efficient frontiers of Value-at-Risk -- return trade-off are plotted in \autoref{fig:VaR_5} for 5\% and \autoref{fig:VaR_10} for 10\% quantile. In both quantiles the model with the best performance is PQR-RV. Similarly to GMVaRP analysis second best performance is achieved by UQR and the model with the worst VaR--return trade--off is RiskMetrics. In \autoref{fig:VaR_10} we can also see that benefits from using PQR are greater for lower values of Value-at-Risk. Overall we can say that Panel Quantile Regression Model for Returns generates better economic performance than the remaining benchmark models.   

\newpage

\section{Conclusion} \label{sec:conclusion}
In this paper, we propose to employ panel quantile regression together with non-parametric measures of quadratic return variation to model conditional quantiles of financial assets return series. For estimation purposes we use penalized fixed effects estimator as introduced in \cite{koenker2004quantile}. Resulting Panel Quantile Regression Model for Returns inherit all favorable properties offered by panel data and quantile regression. A key attraction of the proposed methodology is the ability to control for otherwise unobserved heterogeneity among financial assets so it is possible to disentangle overall market risk into its systemic and idiosyncratic parts. Another attraction is the dimensionality reduction because the number of estimated parameters is always less than or equal to $k+n$, where $k$ is the number of regressors and $n$ number of assets. Last but not least, to the best of our knowledge this is one of the first applications of the panel quantile regression using dataset where $T>>N$. As a result we are able to obtain estimates of quantile specific individual fixed effects that accounts for unobserved heterogeneity and represents idiosyncratic part of the market risk. Moreover these estimates translates into better forecasting performance of newly proposed model compared to traditional benchmarks. Overall we test accuracy and performance of the Panel Quantile Regression Model for Returns in the simple portfolio Value-at-Risk forecasting exercise using simulated and also empirical data. The Monte-Carlo experiment shows that the newly proposed model is dynamically well specified. Moreover when we use heterogeneous data it is able to outperform benchmark models in direct statistical comparison. In the empirical application in-sample model fit highlights importance of the different components of the quadratic variation for the  various quantiles of return series. Out-of-sample statistical comparison shows superiority of the new approach. Better statistical performance moreover translates directly into economic gains as shown by Global Minimum Value-at-Risk Portfolio set-up and efficient frontiers of the Value--at--Risk - Return trade-off. 

Our results make the model attractive not only from academic but also from the practitioners point of view. In particular it is highly attractive for portfolio and risk management because of its ability to handle high-dimensional problems. Importantly it can be easily used to obtain widely used Value-at-Risk measures of portfolios consisting of high number of assets. 

\bibliographystyle{chicago}
\bibliography{Econometrics}
\pagebreak

\section*{Appendix} \label{appendix}
\begin{table}[H]
  \centering
  \caption{Descriptive statistics of daily returns}
	{\def\arraystretch{1.25}\tabcolsep=10pt  
    \begin{tabular}{lcccccc}
    \toprule
          & Mean  & Max   & Min   & St. Dev. & Skewness & Kurtosis \\
\cmidrule{2-7}    AAPL  & -0.05 & 10.62 & -12.29 & 1.72  & -0.14 & 7.09 \\
    AMZN  & 0.09  & 12.32 & -12.96 & 1.95  & 0.33  & 8.27 \\
    BAC   & -0.17 & 19.09 & -25.09 & 2.77  & -0.20 & 20.64 \\
    CMCSA & 0.03  & 12.77 & -13.63 & 1.57  & -0.33 & 12.09 \\
    CSCO  & -0.02 & 7.26  & -8.69 & 1.35  & -0.14 & 7.34 \\
    CVX   & 0.02  & 11.01 & -10.50 & 1.31  & -0.08 & 11.29 \\
    C     & -0.27 & 19.92 & -40.33 & 2.93  & -2.48 & 38.66 \\
    DIS   & 0.06  & 11.03 & -10.29 & 1.36  & 0.34  & 11.11 \\
    GE    & -0.03 & 10.96 & -10.52 & 1.51  & -0.36 & 14.16 \\
    HD    & 0.03  & 11.03 & -7.68 & 1.47  & 0.62  & 9.40 \\
    IBM   & 0.05  & 6.19  & -5.93 & 1.06  & -0.10 & 7.36 \\
    INTC  & 0.01  & 9.20  & -9.43 & 1.42  & 0.13  & 7.41 \\
    JNJ   & 0.01  & 11.19 & -7.77 & 0.85  & 0.75  & 21.90 \\
    JPM   & 0.01  & 13.85 & -19.75 & 2.08  & 0.15  & 16.17 \\
    KO    & 0.02  & 7.14  & -7.37 & 0.93  & -0.08 & 11.52 \\
    MCD   & 0.03  & 8.76  & -7.53 & 1.02  & 0.17  & 9.26 \\
    MRK   & 0.00  & 9.75  & -8.09 & 1.29  & -0.08 & 9.72 \\
    MSFT  & 0.02  & 9.96  & -7.01 & 1.28  & 0.06  & 7.88 \\
    ORCL  & 0.04  & 7.56  & -8.90 & 1.36  & -0.09 & 6.85 \\
    PEP   & 0.04  & 8.44  & -6.27 & 0.90  & 0.32  & 10.24 \\
    PFE   & -0.03 & 6.49  & -7.46 & 1.14  & -0.07 & 7.02 \\
    PG    & 0.05  & 7.07  & -5.62 & 0.86  & 0.00  & 9.50 \\
    QCOM  & -0.01 & 9.04  & -8.15 & 1.45  & -0.10 & 6.31 \\
    SLB   & 0.00  & 11.34 & -15.62 & 1.85  & -0.33 & 9.57 \\
    T     & -0.01 & 11.42 & -6.56 & 1.11  & 0.50  & 13.58 \\
    VZ    & 0.01  & 8.62  & -7.72 & 1.12  & 0.40  & 10.41 \\
    WFC   & 0.00  & 18.29 & -18.73 & 2.23  & 0.45  & 18.50 \\
    WMT   & 0.00  & 7.71  & -10.60 & 0.97  & -0.08 & 14.66 \\
    XOM   & 0.03  & 8.90  & -11.76 & 1.22  & -0.11 & 13.33 \\
    \bottomrule
    \end{tabular}%
    }
    \begin{tablenotes}
\footnotesize
\item{Note: Values for Mean, Max, Min and St. Dev are displayed in \%.}
\end{tablenotes}
  \label{tab:descriptive_stat}%
\end{table}%

\begin{table}[H]
\begin{center}
\small
\caption{Univariate Normal Distribution -- Mean of coefficients estimates from Monte-Carlo simulations} \label{tab:sim_UND_insample_PQR}
\begin{tabular}{lccccccc}
\toprule
$\tau$ & 5\% & 10\% & 25\% & 50\% & 75\% & 90\% & 95\% \\ \cmidrule{2-8} \\[-0.75em]
 & \multicolumn{7}{c}{\textit{PQR-RV}} \\ \cmidrule{2-8}
$\hat{\beta}_{RV^{1/2}}$ & -1.56 & -1.22 & -0.64 & -0.01 & 0.62 & 1.2 & 1.54 \\ 
 & (-44.19) & (-43.75) & (-30.85) & (-0.6) & (30.86) & (44.63) & (46.36) \\   \cmidrule{2-8} \\[-0.75em]
 & \multicolumn{7}{c}{\textit{PQR-RSV}} \\ \cmidrule{2-8} 
$\hat{\beta}_{{RS^{+}}^{1/2}}$ & -1.12 & -0.87 & -0.46 & -0.01 & 0.45 & 0.84 & 1.07 \\ 
 & (-5.38) & (-6.06) & (-5.34) & (-0.12) & (5.55) & (5.66) & (5.28) \\[0.5em]
$\hat{\beta}_{{RS^{-}}^{1/2}}$ & -1.09 & -0.86 & -0.46 & -0.01 & 0.45 & 0.86 & 1.11 \\ 
 & (-5.29) & (-6.07) & (-5.44) & (-0.18) & (5.58) & (5.78) & (5.5) \\ \cmidrule{2-8} \\[-0.75em]
 & \multicolumn{7}{c}{\textit{PQR-BPV}} \\ \cmidrule{2-8} 
$\hat{\beta}_{BPV^{1/2}}$ & -1.58 & -1.25 & -0.67 & -0.01 & 0.65 & 1.23 & 1.56 \\ 
 & (-45.75) & (-45.44) & (-31.71) & (-0.6) & (32.26) & (46.06) & (47.31) \\[0.5em]
$\hat{\beta}_{Jumps^{1/2}}$ & 0.08 & 0.06 & 0.03 & 0 & -0.03 & -0.06 & -0.08 \\ 
 & (1.1) & (1.14) & (0.74) & (-0.03) & (-0.83) & (-1.1) & (-1.05) \\ 
\bottomrule
\end{tabular}
\begin{tablenotes}
\centering
\footnotesize
\item{Note: Table displays mean of coefficient estimates with corresponding t-statistics in parentheses. Individual fixed effects $\alpha_i(\tau)$ are not reported for brevity.}
\end{tablenotes}
\end{center}
\end{table}

\begin{table}[H]
\begin{center}
\small
\caption{Multivariate Student-t Distribution -- Mean of coefficients estimates from Monte-Carlo simulations} \label{tab:sim_MTD_insample_PQR}
\begin{tabular}{lccccccc}
\toprule
$\tau$ & 5\% & 10\% & 25\% & 50\% & 75\% & 90\% & 95\% \\ \cmidrule{2-8} \\[-0.75em]
 & \multicolumn{7}{c}{\textit{PQR-RV}} \\ \cmidrule{2-8}
$\hat{\beta}_{RV^{1/2}}$ & -1.56 & -1.16 & -0.58 & -0.01 & 0.57 & 1.15 & 1.55 \\ 
 & (-18.43) & (-19.14) & (-13.3) & (-0.17) & (12.9) & (18.73) & (19.43) \\   \cmidrule{2-8} \\[-0.75em]
 & \multicolumn{7}{c}{\textit{PQR-RSV}} \\ \cmidrule{2-8} 
$\hat{\beta}_{{RS^{+}}^{1/2}}$ & -1.12 & -0.83 & -0.42 & -0.01 & 0.39 & 0.82 & 1.09 \\ 
 & (-2.55) & (-2.66) & (-2.2) & (-0.12) & (2.05) & (2.55) & (2.43) \\[0.5em]
$\hat{\beta}_{{RS^{-}}^{1/2}}$ & -1.09 & -0.82 & -0.4 & 0 & 0.41 & 0.81 & 1.1 \\ 
 & (-2.48) & (-2.62) & (-2.13) & (0.04) & (2.15) & (2.52) & (2.42) \\ \cmidrule{2-8} \\[-0.75em]
 & \multicolumn{7}{c}{\textit{PQR-BPV}} \\ \cmidrule{2-8} 
$\hat{\beta}_{BPV^{1/2}}$ & -1.62 & -1.21 & -0.6 & -0.01 & 0.59 & 1.19 & 1.6 \\ 
 & (-18.27) & (-19) & (-13.14) & (-0.17) & (12.64) & (18.41) & (19.05) \\[0.5em]
$\hat{\beta}_{Jumps^{1/2}}$ & -0.06 & -0.04 & -0.02 & 0 & 0.02 & 0.05 & 0.06 \\ 
 & (-0.45) & (-0.45) & (-0.29) & (0.04) & (0.29) & (0.47) & (0.45) \\ 
\bottomrule
\end{tabular}
\begin{tablenotes}
\centering
\footnotesize
\item{Note: Table displays mean of coefficient estimates with corresponding t-statistics in parentheses. Individual fixed effects $\alpha_i(\tau)$ are not reported for brevity.}
\end{tablenotes}
\end{center}
\end{table}
\pagebreak

\begin{table}[H]
\begin{center}
\small
\caption{Univariate Student-t Distribution -- Mean of coefficients estimates from Monte-Carlo simulations} \label{tab:sim_UTD_insample_PQR}
\begin{tabular}{lccccccc}
\toprule
$\tau$ & 5\% & 10\% & 25\% & 50\% & 75\% & 90\% & 95\% \\ \cmidrule{2-8} \\[-0.75em]
 & \multicolumn{7}{c}{\textit{PQR-RV}} \\ \cmidrule{2-8}
$\hat{\beta}_{RV^{1/2}}$ & -1.56 & -1.23 & -0.65 & -0.01 & 0.63 & 1.21 & 1.54 \\ 
 & (-52.18) & (-52.37) & (-36.12) & (-0.53) & (35.09) & (55.55) & (52.99) \\     \cmidrule{2-8} \\[-0.75em]
 & \multicolumn{7}{c}{\textit{PQR-RSV}} \\ \cmidrule{2-8} 
$\hat{\beta}_{{RS^{+}}^{1/2}}$ & -1.12 & -0.89 & -0.47 & -0.01 & 0.44 & 0.85 & 1.08 \\ 
 & (-6.47) & (-6.84) & (-5.75) & (-0.25) & (5.48) & (6.39) & (6.13) \\[0.5em]
$\hat{\beta}_{{RS^{-}}^{1/2}}$ & -1.09 & -0.86 & -0.45 & 0 & 0.46 & 0.87 & 1.12 \\ 
 & (-6.23) & (-6.57) & (-5.65) & (0.01) & (5.84) & (6.55) & (6.39) \\ \cmidrule{2-8} \\[-0.75em]
 & \multicolumn{7}{c}{\textit{PQR-BPV}} \\ \cmidrule{2-8} 
$\hat{\beta}_{BPV^{1/2}}$ & -1.64 & -1.29 & -0.69 & -0.01 & 0.67 & 1.27 & 1.62 \\ 
 & (-51.41) & (-52.71) & (-36.1) & (-0.52) & (35.3) & (55.28) & (52.55) \\[0.5em]
$\hat{\beta}_{Jumps^{1/2}}$ & -0.02 & -0.01 & 0 & 0 & 0 & 0.01 & 0.02 \\ 
 & (-0.3) & (-0.21) & (-0.11) & (-0.02) & (0.01) & (0.16) & (0.24) \\
\bottomrule
\end{tabular}
\begin{tablenotes}
\centering
\footnotesize
\item{Note: Table displays mean of coefficient estimates with corresponding t-statistics in parentheses. Individual fixed effects $\alpha_i(\tau)$ are not reported for brevity.}
\end{tablenotes}
\end{center}
\end{table}
\pagebreak

\begin{sidewaystable}[H]
\centering
\captionof{table}{Models performance using data simulated from Multivariate Student-t Distribution}
\resizebox{\textwidth}{!}{
\begin{tabular}{lcccccccccccccccccc}
    \toprule
          &       & \multicolumn{5}{c}{PQR-RV}            &       & \multicolumn{5}{c}{PQR-RSV}           &       & \multicolumn{5}{c}{PQR-BPV} \\
\cmidrule{3-7}\cmidrule{9-13}\cmidrule{15-19}    \textit{Panel A.1} &       & 5\%   & 10\%  & 50\%  & 90\%  & 95\%  &       & 5\%   & 10\%  & 50\%  & 90\%  & 95\%  &       & 5\%   & 10\%  & 50\%  & 90\%  & 95\% \\
    \midrule
          & $\widehat{DQ}_{violations}$ & 6.4   & 5.8   & 11    & 5.6   & 7.8   &       & 6     & 6.2   & 11.4  & 5.2   & 7     &       & 6.8   & 5.8   & 10.6  & 5.2   & 7.6 \\
          & $\widehat{\tau}_{avg}$ & 5.1   & 10.2  & 51.3  & 90.0  & 95.0  &       & 5.1   & 10.2  & 51.4  & 90.0  & 95.0  &       & 5.1   & 10.2  & 51.4  & 90.0  & 95.0 \\
          & $\widehat{\tau}_{max}$ & 6.5   & 12.4  & 55.3  & 91.7  & 96.3  &       & 6.5   & 12.4  & 55.1  & 91.7  & 96.2  &       & 6.5   & 12.5  & 55.3  & 91.8  & 96.3 \\
          & $\widehat{\tau}_{min}$ & 3.8   & 7.9   & 48.1  & 87.8  & 93.7  &       & 3.7   & 7.9   & 48.2  & 88.0  & 93.8  &       & 3.8   & 7.9   & 48.2  & 88.1  & 93.7 \\
          & $\widehat{\tau}_{avg-dev}$ & 0.1   & 0.2   & 1.3   & 0.0   & 0.0   &       & 0.1   & 0.2   & 1.4   & 0.0   & 0.0   &       & 0.1   & 0.2   & 1.4   & -0.1  & 0.0 \\
          &       &       &       &       &       &       &       &       &       &       &       &       &       &       &       &       &       &  \\
          &       & \multicolumn{5}{c}{RiskMetrics}       &       & \multicolumn{5}{c}{UQR}               &       & \multicolumn{5}{c}{Portfolio UQR} \\
\cmidrule{3-7}\cmidrule{9-13}\cmidrule{15-19}    \textit{Panel A.2} &       & 5\%   & 10\%  & 50\%  & 90\%  & 95\%  &       & 5\%   & 10\%  & 50\%  & 90\%  & 95\%  &       & 5\%   & 10\%  & 50\%  & 90\%  & 95\% \\
    \midrule
          & $\widehat{DQ}_{violations}$ & 5.6   & 12.4  & 5.6   & 18.6  & 8.2   &       & 6.4   & 5     & 13.4  & 6.2   & 7.6   &       & 5.8   & 6.2   & 4.2   & 6.8   & 6.6 \\
          & $\widehat{\tau}_{avg}$ & 5.2   & 9.2   & 50.4  & 91.0  & 94.9  &       & 5.0   & 10.2  & 51.5  & 90.0  & 95.0  &       & 4.8   & 9.6   & 50.0  & 90.4  & 95.2 \\
          & $\widehat{\tau}_{max}$ & 6.8   & 11.7  & 54.0  & 92.8  & 96.3  &       & 6.4   & 12.3  & 55.2  & 91.7  & 96.5  &       & 6.1   & 11.2  & 53.1  & 92.2  & 96.3 \\
          & $\widehat{\tau}_{min}$ & 4.0   & 7.2   & 46.2  & 89.1  & 93.4  &       & 3.7   & 7.9   & 48.0  & 87.9  & 93.9  &       & 3.7   & 7.8   & 46.8  & 88.1  & 93.7 \\
          & $\widehat{\tau}_{avg-dev}$ & 0.2   & -0.8  & 0.4   & 1.0   & -0.1  &       & 0.0   & 0.2   & 1.5   & 0.0   & 0.0   &       & -0.2  & -0.4  & 0.0   & 0.4   & 0.2 \\
          &       &       &       &       &       &       &       &       &       &       &       &       &       &       &       &       &       &  \\
          &       & \multicolumn{17}{c}{benchmark} \\
\cmidrule{3-19}          &       & \multicolumn{5}{c}{RiskMetrics}       &       & \multicolumn{5}{c}{UQR}               &       & \multicolumn{5}{c}{Portfolio UQR} \\
\cmidrule{3-7}\cmidrule{9-13}\cmidrule{15-19}    \textit{Panel B.1} &       & 5\%   & 10\%  & 50\%  & 90\%  & 95\%  &       & 5\%   & 10\%  & 50\%  & 90\%  & 95\%  &       & 5\%   & 10\%  & 50\%  & 90\%  & 95\% \\
    \midrule
    PQR-RV & $DM$  & 54    & 58.2  & 0.2   & 62.6  & 57.8  &       & 2.4   & 2.4   & 12    & 3.6   & 3.6   &       & 43    & 40    & 18.4  & 46.4  & 49.2 \\
    PQR-RSV & $DM$  & 54.2  & 57.4  & 0.2   & 62.2  & 57.4  &       & 1.4   & 1.2   & 9.6   & 1.4   & 3     &       & 41    & 39    & 19.2  & 45    & 48.2 \\
    PQR-BPV & $DM$  & 53    & 57.6  & 0.2   & 62.4  & 57.4  &       & 1.6   & 2     & 10    & 3     & 5.2   &       & 42.4  & 38.6  & 18.8  & 44.8  & 45.4 \\
          &       &       &       &       &       &       &       &       &       &       &       &       &       &       &       &       &       &  \\
          &       & \multicolumn{5}{c}{PQR-RV}            &       & \multicolumn{5}{c}{PQR-RSV}           &       & \multicolumn{5}{c}{PQR-BPV} \\
\cmidrule{3-7}\cmidrule{9-13}\cmidrule{15-19}    \textit{Panel B.2} &       & 5\%   & 10\%  & 50\%  & 90\%  & 95\%  &       & 5\%   & 10\%  & 50\%  & 90\%  & 95\%  &       & 5\%   & 10\%  & 50\%  & 90\%  & 95\% \\
    \midrule
    RiskMetrics & $DM$  & 0     & 0     & 14.4  & 0     & 0     &       & 0     & 0     & 14.8  & 0     & 0     &       & 0     & 0     & 14.6  & 0     & 0 \\
    UQR & $DM$  & 9     & 10.6  & 2.8   & 9.8   & 10.6  &       & 10.4  & 11    & 2.8   & 11.2  & 9.6   &       & 10.4  & 11.6  & 2.8   & 8.2   & 7.8 \\
    Portfolio UQR & $DM$  & 0.8   & 0.4   & 1     & 0.6   & 0.4   &       & 1     & 0.4   & 1.4   & 0.6   & 0.4   &       & 0.8   & 0.6   & 1.2   & 0.8   & 0.2 \\
    \bottomrule
    \end{tabular}%
    }
\begin{tablenotes}
\footnotesize
\item{Note: Table displays absolute and relative performance of PQR models for returns with RV, RSV and BPV as regressors and benchmark models.}
\medskip
\item {\textit{Panel A.1} reports absolute performance of PQR models, \textit{Panel A.2} reports absolute performance of  benchmark models. For each model and quantile $\tau$, percentage of violations of the CAViaR test for correct dynamic specification ($\widehat{DQ}_{violations}$), average unconditional coverage ($\widehat{\tau}_{avg}$),maximum unconditional coverage ($\widehat{\tau}_{max}$), minimum unconditional coverage ($\widehat{\tau}_{min}$) and average deviation of unconditional coverage from given quantile $\tau$ ($\widehat{\tau}_{avg-dev}$)}
\medskip
\item{\textit{Panel B.1} and \textit{Panel B.2} report relative performance of Panel Quantile Regression Models for Returns in comparison to benchmark models and relative performance of benchmark models in comparison to Panel Quantile Regression Models for Returns respectively. For each specification and quantile $\tau$ we report percentage of statistically better performance according to Diebold-Mariano($DM$) test at 5\% significance level.}
\end{tablenotes}
  \label{tab:sim MTD}%
\end{sidewaystable}

\begin{sidewaystable}[H]
\centering
\captionof{table}{Models performance using data simulated from Univariate Normal Distributions}
\resizebox{\textwidth}{!}{
    \begin{tabular}{lcccccccccccccccccc}
    \toprule
          &       & \multicolumn{5}{c}{PQR-RV}            &       & \multicolumn{5}{c}{PQR-RSV}           &       & \multicolumn{5}{c}{PQR-BPV} \\
\cmidrule{3-7}\cmidrule{9-13}\cmidrule{15-19}    \textit{Panel A.1} &       & 5\%   & 10\%  & 50\%  & 90\%  & 95\%  &       & 5\%   & 10\%  & 50\%  & 90\%  & 95\%  &       & 5\%   & 10\%  & 50\%  & 90\%  & 95\% \\
    \midrule
          & $\widehat{DQ}_{violations}$ & 8.2   & 7.4   & 26    & 4.4   & 8.6   &       & 9     & 8.4   & 26.6  & 4.6   & 7.4   &       & 9.4   & 9.6   & 26.4  & 4.2   & 6.8 \\
          & $\widehat{\tau}_{avg}$ & 5.3   & 10.6  & 52.5  & 90.3  & 95.2  &       & 5.4   & 10.6  & 52.5  & 90.3  & 95.2  &       & 5.4   & 10.7  & 52.5  & 90.3  & 95.2 \\
          & $\widehat{\tau}_{max}$ & 7.1   & 12.7  & 55.6  & 92.9  & 96.9  &       & 7.1   & 12.6  & 55.7  & 92.8  & 96.8  &       & 7.2   & 12.8  & 55.7  & 92.7  & 96.8 \\
          & $\widehat{\tau}_{min}$ & 3.8   & 8.8   & 49.3  & 88.1  & 93.5  &       & 3.9   & 8.8   & 49.3  & 88.1  & 93.4  &       & 3.8   & 8.6   & 49.3  & 88.1  & 93.5 \\
          & $\widehat{\tau}_{avg-dev}$ & 0.3   & 0.6   & 2.5   & 0.3   & 0.2   &       & 0.4   & 0.6   & 2.5   & 0.3   & 0.2   &       & 0.4   & 0.7   & 2.5   & 0.3   & 0.2 \\
          &       &       &       &       &       &       &       &       &       &       &       &       &       &       &       &       &       &  \\
          &       & \multicolumn{5}{c}{RiskMetrics}       &       & \multicolumn{5}{c}{UQR}               &       & \multicolumn{5}{c}{Portfolio UQR} \\
\cmidrule{3-7}\cmidrule{9-13}\cmidrule{15-19}    \textit{Panel A.2} &       & 5\%   & 10\%  & 50\%  & 90\%  & 95\%  &       & 5\%   & 10\%  & 50\%  & 90\%  & 95\%  &       & 5\%   & 10\%  & 50\%  & 90\%  & 95\% \\
    \midrule
          & $\widehat{DQ}_{violations}$ & 18.8  & 13.4  & 10    & 5.4   & 9.8   &       & 7.4   & 8.2   & 47.6  & 4.2   & 7.4   &       & 5.4   & 4.4   & 3     & 3     & 4.4 \\
          & $\widehat{\tau}_{avg}$ & 5.8   & 11.0  & 51.2  & 90.2  & 94.9  &       & 5.3   & 10.6  & 53.3  & 90.3  & 95.2  &       & 5.1   & 10.1  & 50.0  & 89.9  & 94.9 \\
          & $\widehat{\tau}_{max}$ & 7.3   & 12.7  & 54.1  & 92.1  & 96.2  &       & 7.1   & 12.6  & 56.4  & 92.9  & 96.9  &       & 6.3   & 11.6  & 52.3  & 91.7  & 96.1 \\
          & $\widehat{\tau}_{min}$ & 4.3   & 9.1   & 47.8  & 88.0  & 93.4  &       & 3.8   & 8.7   & 50.2  & 88.3  & 93.4  &       & 4.1   & 8.4   & 47.7  & 88.2  & 93.7 \\
          & $\widehat{\tau}_{avg-dev}$ & 0.8   & 1.0   & 1.2   & 0.2   & -0.1  &       & 0.3   & 0.6   & 3.3   & 0.3   & 0.2   &       & 0.1   & 0.1   & 0.0   & -0.1  & -0.1 \\
          &       &       &       &       &       &       &       &       &       &       &       &       &       &       &       &       &       &  \\
          &       & \multicolumn{17}{c}{benchmark} \\
\cmidrule{3-19}          &       & \multicolumn{5}{c}{RiskMetrics}       &       & \multicolumn{5}{c}{UQR}               &       & \multicolumn{5}{c}{Portfolio UQR} \\
\cmidrule{3-7}\cmidrule{9-13}\cmidrule{15-19}    \textit{Panel B.1} &       & 5\%   & 10\%  & 50\%  & 90\%  & 95\%  &       & 5\%   & 10\%  & 50\%  & 90\%  & 95\%  &       & 5\%   & 10\%  & 50\%  & 90\%  & 95\% \\
    \midrule
    PQR-RV & $DM$  & 62.6  & 56.6  & 0     & 57    & 60.6  &       & 9     & 6     & 72.4  & 6.6   & 7.4   &       & 16.4  & 17.4  & 4.6   & 17.4  & 16.4 \\
    PQR-RSV & $DM$  & 63.2  & 57.6  & 0     & 56.8  & 61.8  &       & 12    & 7     & 71.8  & 9.2   & 12.4  &       & 18.4  & 18.8  & 4.6   & 17.2  & 17.4 \\
    PQR-BPV & $DM$  & 65.4  & 58.8  & 0     & 58.6  & 62.2  &       & 12.8  & 11.2  & 71.4  & 12.4  & 14.4  &       & 19.2  & 18.6  & 4.6   & 19.4  & 16.4 \\
          &       &       &       &       &       &       &       &       &       &       &       &       &       &       &       &       &       &  \\
          &       & \multicolumn{5}{c}{PQR-RV}            &       & \multicolumn{5}{c}{PQR-RSV}           &       & \multicolumn{5}{c}{PQR-BPV} \\
\cmidrule{3-7}\cmidrule{9-13}\cmidrule{15-19}    \textit{Panel B.2} &       & 5\%   & 10\%  & 50\%  & 90\%  & 95\%  &       & 5\%   & 10\%  & 50\%  & 90\%  & 95\%  &       & 5\%   & 10\%  & 50\%  & 90\%  & 95\% \\
    \midrule
    RiskMetrics & $DM$  & 0     & 0     & 44.2  & 0     & 0     &       & 0     & 0     & 44.2  & 0     & 0     &       & 0     & 0     & 44.4  & 0.2   & 0 \\
    UQR   & $DM$  & 2.8   & 2.8   & 0     & 2.8   & 2.8   &       & 2.2   & 2     & 0     & 2.2   & 1.8   &       & 2.4   & 1.8   & 0     & 3     & 1.2 \\
    Portfolio UQR & $DM$  & 0.2   & 0.4   & 2.4   & 0.6   & 1     &       & 0.2   & 0.4   & 2.4   & 0.6   & 0.8   &       & 0.2   & 0.2   & 2.4   & 1.2   & 0.4 \\
    \bottomrule
    \end{tabular}%
    }
\begin{tablenotes}
\footnotesize
\item{Note: Table displays absolute and relative performance of PQR models for returns with RV, RSV and BPV as regressors and benchmark models.}
\medskip
\item {\textit{Panel A.1} reports absolute performance of PQR models, \textit{Panel A.2} reports absolute performance of  benchmark models. For each model and quantile $\tau$, percentage of violations of the CAViaR test for correct dynamic specification ($\widehat{DQ}_{violations}$), average unconditional coverage ($\widehat{\tau}_{avg}$),maximum unconditional coverage ($\widehat{\tau}_{max}$), minimum unconditional coverage ($\widehat{\tau}_{min}$) and average deviation of unconditional coverage from given quantile $\tau$ ($\widehat{\tau}_{avg-dev}$)}
\medskip
\item{\textit{Panel B.1} and \textit{Panel B.2} report relative performance of Panel Quantile Regression Models for Returns in comparison to benchmark models and relative performance of benchmark models in comparison to Panel Quantile Regression Models for Returns respectively. For each specification and quantile $\tau$ we report percentage of statistically better performance according to Diebold-Mariano($DM$) test at 5\% significance level.}
\end{tablenotes}
  \label{tab:sim UND}%
\end{sidewaystable}

\begin{sidewaystable}[H]
\centering
\captionof{table}{Models performance using data simulated from Univariate Student-t Distributions}
\resizebox{\textwidth}{!}{
    \begin{tabular}{lcccccccccccccccccc}
    \toprule
          &       & \multicolumn{5}{c}{PQR-RV}            &       & \multicolumn{5}{c}{PQR-RSV}           &       & \multicolumn{5}{c}{PQR-BPV} \\
\cmidrule{3-7}\cmidrule{9-13}\cmidrule{15-19}    \textit{Panel A.1} &       & 5\%   & 10\%  & 50\%  & 90\%  & 95\%  &       & 5\%   & 10\%  & 50\%  & 90\%  & 95\%  &       & 5\%   & 10\%  & 50\%  & 90\%  & 95\% \\
    \midrule
          & $\widehat{DQ}_{violations}$ & 7.2   & 8.0   & 23.4  & 4.0   & 6.4   &       & 7.0   & 8.2   & 23.2  & 4.0   & 6.6   &       & 7.6   & 8.2   & 23.6  & 4.2   & 6.8 \\
          & $\widehat{\tau}_{avg}$ & 5.3   & 10.5  & 52.3  & 90.2  & 95.2  &       & 5.3   & 10.5  & 52.3  & 90.2  & 95.2  &       & \textbf{5.3} & 10.5  & 52.4  & 90.2  & 95.1 \\
          & $\widehat{\tau}_{max}$ & 7.0   & 12.8  & 56.0  & 92.0  & 96.6  &       & 7.0   & 12.8  & 56.0  & 91.9  & 96.6  &       & 7.0   & 12.7  & 56.1  & 91.9  & 96.6 \\
          & $\widehat{\tau}_{min}$ & 3.9   & 8.3   & 47.7  & 87.8  & 93.6  &       & 3.8   & 8.2   & 47.7  & 87.8  & 93.6  &       & 4.0   & 8.1   & 47.8  & 87.8  & 93.4 \\
          & $\widehat{\tau}_{avg-dev}$ & 0.3   & 0.5   & 2.3   & 0.2   & 0.2   &       & 0.3   & 0.5   & 2.3   & 0.2   & 0.2   &       & 0.3   & 0.5   & 2.4   & 0.2   & 0.1 \\
          &       &       &       &       &       &       &       &       &       &       &       &       &       &       &       &       &       &  \\
          &       & \multicolumn{5}{c}{RiskMetrics}       &       & \multicolumn{5}{c}{UQR}               &       & \multicolumn{5}{c}{Portfolio UQR} \\
\cmidrule{3-7}\cmidrule{9-13}\cmidrule{15-19}    \textit{Panel A.2} &       & 5\%   & 10\%  & 50\%  & 90\%  & 95\%  &       & 5\%   & 10\%  & 50\%  & 90\%  & 95\%  &       & 5\%   & 10\%  & 50\%  & 90\%  & 95\% \\
    \midrule
          & $\widehat{DQ}_{violations}$ & 14.8  & 10.0  & 7.6   & 5.6   & 6.4   &       & 6.8   & 7.4   & 46.4  & 5.0   & 7.0   &       & 3.4   & 4.2   & 2.6   & 3.8   & 3.2 \\
          & $\widehat{\tau}_{avg}$ & 5.8   & 10.9  & 51.2  & 90.1  & 94.8  &       & 5.3   & 10.5  & 53.2  & 90.2  & 95.2  &       & 5.1   & 10.1  & 50.0  & 89.9  & 94.9 \\
          & $\widehat{\tau}_{max}$ & 7.3   & 12.8  & 55.0  & 91.7  & 96.3  &       & 7.0   & 12.6  & 56.9  & 91.9  & 96.5  &       & 6.8   & 12.5  & 52.2  & 91.2  & 96.0 \\
          & $\widehat{\tau}_{min}$ & 4.3   & 8.9   & 46.8  & 88.1  & 93.4  &       & 3.9   & 8.3   & 48.5  & 87.9  & 93.5  &       & 4.0   & 8.2   & 47.4  & 88.0  & 93.7 \\
          & $\widehat{\tau}_{avg-dev}$ & 0.8   & 0.9   & 1.2   & 0.1   & -0.2  &       & 0.3   & 0.5   & 3.2   & 0.2   & 0.2   &       & 0.1   & 0.1   & 0.0   & -0.1  & -0.1 \\
          &       &       &       &       &       &       &       &       &       &       &       &       &       &       &       &       &       &  \\
          &       & \multicolumn{17}{c}{benchmark} \\
\cmidrule{3-19}          &       & \multicolumn{5}{c}{RiskMetrics}       &       & \multicolumn{5}{c}{UQR}               &       & \multicolumn{5}{c}{Portfolio UQR} \\
\cmidrule{3-7}\cmidrule{9-13}\cmidrule{15-19}    \textit{Panel B.1} &       & 5\%   & 10\%  & 50\%  & 90\%  & 95\%  &       & 5\%   & 10\%  & 50\%  & 90\%  & 95\%  &       & 5\%   & 10\%  & 50\%  & 90\%  & 95\% \\
    \midrule
    PQR-RV & $DM$  & 66.4  & 57.8  & 0.4   & 57.4  & 59.4  &       & 6.4   & 6.4   & 67.4  & 9.2   & 7.8   &       & 18.6  & 17.0  & 7.6   & 23.2  & 20.4 \\
    PQR-RSV & $DM$  & 65.8  & 58.2  & 0.4   & 58.0  & 59.6  &       & 6.8   & 7.8   & 68.6  & 10.2  & 8.2   &       & 17.2  & 17.4  & 7.4   & 23.2  & 21.6 \\
    PQR-BPV & $DM$  & 67.0  & 59.6  & 0.4   & 59.4  & 61.2  &       & 9.0   & 6.8   & 66.8  & 10.6  & 10.8  &       & 20.6  & 16.6  & 7.6   & 22.4  & 21.6 \\
          &       &       &       &       &       &       &       &       &       &       &       &       &       &       &       &       &       &  \\
          &       & \multicolumn{5}{c}{PQR-RV}            &       & \multicolumn{5}{c}{PQR-RSV}           &       & \multicolumn{5}{c}{PQR-BPV} \\
\cmidrule{3-7}\cmidrule{9-13}\cmidrule{15-19}    \textit{Panel B.2} &       & 5\%   & 10\%  & 50\%  & 90\%  & 95\%  &       & 5\%   & 10\%  & 50\%  & 90\%  & 95\%  &       & 5\%   & 10\%  & 50\%  & 90\%  & 95\% \\
    \midrule
    RiskMetrics & $DM$  & 0.0   & 0.0   & 39.6  & 0.0   & 0.0   &       & 0.0   & 0.0   & 39.8  & 0.0   & 0.0   &       & 0.0   & 0.0   & 39.6  & 0.0   & 0.0 \\
    UQR   & $DM$  & 4.0   & 4.0   & 0.0   & 4.0   & 4.4   &       & 4.6   & 3.2   & 0.2   & 3.8   & 3.6   &       & 3.4   & 2.4   & 0.0   & 2.6   & 3.0 \\
    Portfolio UQR & $DM$  & 0.2   & 0.0   & 2.6   & 0.0   & 0.4   &       & 0.2   & 0.0   & 2.8   & 0.0   & 0.2   &       & 0.2   & 0.0   & 2.8   & 0.0   & 0.4 \\
    \bottomrule
    \end{tabular}%
    }
\begin{tablenotes}
\footnotesize
\item{Note: Table displays absolute and relative performance of PQR models for returns with RV, RSV and BPV as regressors and benchmark models.}
\medskip
\item {\textit{Panel A.1} reports absolute performance of PQR models, \textit{Panel A.2} reports absolute performance of  benchmark models. For each model and quantile $\tau$, percentage of violations of the CAViaR test for correct dynamic specification ($\widehat{DQ}_{violations}$), average unconditional coverage ($\widehat{\tau}_{avg}$),maximum unconditional coverage ($\widehat{\tau}_{max}$), minimum unconditional coverage ($\widehat{\tau}_{min}$) and average deviation of unconditional coverage from given quantile $\tau$ ($\widehat{\tau}_{avg-dev}$)}
\medskip
\item{\textit{Panel B.1} and \textit{Panel B.2} report relative performance of Panel Quantile Regression Models for Returns in comparison to benchmark models and relative performance of benchmark models in comparison to Panel Quantile Regression Models for Returns respectively. For each specification and quantile $\tau$ we report percentage of statistically better performance according to Diebold-Mariano($DM$) test at 5\% significance level.}
\end{tablenotes}
  \label{tab:sim UND}%
\end{sidewaystable}

\begin{table}[H]
\begin{center}
\small
\caption{Coefficient estimates of Panel Quantile Regressions: $\lambda=1$} \label{tab:insample_PQR_lambda_1}
\begin{tabular}{lccccccc}
\toprule
$\tau$ & 5\% & 10\% & 25\% & 50\% & 75\% & 90\% & 95\% \\ \cmidrule{2-8} \\[-0.75em]
 & \multicolumn{7}{c}{\textit{PQR-RV}} \\ \cmidrule{2-8}
$const$ & 0 & 0 & 0 & 0 & 0 & 0 & 0 \\
 & (-1.47) & (-1.27) & (-0.61) & (-0.58) & (0.9) & (1.64) & (2.08) \\[0.5em]
$\hat{\beta}_{RV^{1/2}}$ & -1.51 & -1.16 & -0.6 & -0.01 & 0.56 & 1.11 & 1.42 \\
 & (-24.24) & (-21.41) & (-16.36) & (-0.24) & (20.15) & (24.62) & (21.11) \\ \cmidrule{2-8} \\[-0.75em]
 & \multicolumn{7}{c}{\textit{PQR-RSV}} \\ \cmidrule{2-8} 
$const$ & 0 & 0 & 0 & 0 & 0 & 0 & 0 \\
 & (-1.6) & (-1.21) & (-0.63) & (-0.44) & (0.97) & (2.07) & (2.41) \\[0.5em]
$\hat{\beta}_{{RS^{+}}^{1/2}}$ & -0.97 & -0.74 & -0.44 & -0.15 & 0.18 & 0.41 & 0.54 \\
 & (-12.92) & (-13.02) & (-8.54) & (-2.9) & (2.82) & (4.39) & (4.3) \\[0.5em]
$\hat{\beta}_{{RS^{-}}^{1/2}}$ & -1.18 & -0.91 & -0.41 & 0.14 & 0.62 & 1.15 & 1.49 \\
 & (-11.14) & (-14.29) & (-10.12) & (2.78) & (9.23) & (13.91) & (10.06) \\ \cmidrule{2-8} \\[-0.75em]
 & \multicolumn{7}{c}{\textit{PQR-BPV}} \\ \cmidrule{2-8} 
$const$ & 0 & 0 & 0 & 0 & 0 & 0 & 0 \\
 & (-1.4) & (-1.36) & (-0.6) & (-0.67) & (0.77) & (1.86) & (2.67) \\[0.5em]
$\hat{\beta}_{BPV^{1/2}}$ & -1.55 & -1.18 & -0.62 & 0 & 0.59 & 1.15 & 1.44 \\
 & (-20.25) & (-17.58) & (-16.49) & (-0.15) & (24.16) & (22.79) & (25.91) \\[0.5em]
$\hat{\beta}_{Jumps^{1/2}}$ & -0.24 & -0.2 & -0.14 & -0.03 & 0.06 & 0.21 & 0.44 \\
 & (-3.2) & (-3.41) & (-3.41) & (-0.62) & (1.03) & (1.88) & (2.73) \\ 
\bottomrule
\end{tabular}
\begin{tablenotes}
\centering
\footnotesize
\item{Note: Table displays coefficient estimates with bootstraped t-statistics in parentheses. Individual fixed effects $\alpha_i(\tau)$ are not reported for brevity - they are available from authors upon request}
\end{tablenotes}
\end{center}
\end{table}

\end{document}